\documentclass[12pt]{iopart}
\pdfoutput=1
\usepackage{iopams}
\usepackage{graphicx}
\usepackage{amssymb,amsthm}

\def\im{\mathop{\rm Im}\nolimits}

\begin{document}
\title{Partition functions and the continuum limit in Penner matrix models}
\author{ Gabriel \'Alvarez$^1$, Luis Mart\'{\i}nez Alonso$^1$ and Elena Medina$^2$}
\address{$^1$ Departamento de F\'{\i}sica Te\'orica II,
                        Facultad de Ciencias F\'{\i}sicas,
                        Universidad Complutense,
                        28040 Madrid, Spain}
\address{$^2$ Departamento de Matem\'aticas,
                        Facultad de Ciencias,
                        Universidad de C\'adiz,
                        11510 Puerto Real, C\'adiz, Spain}
\begin{abstract}
We present an implementation of the method of orthogonal polynomials
which  is particularly suitable to study the partition functions of Penner random matrix models,
to obtain their explicit forms in the exactly solvable cases, and to determine
the coefficients of  their perturbative expansions in the continuum limit.
The method relies on identities satisfied by the resolvent of the Jacobi matrix
in the three-term recursion relation of the associated families of orthogonal polynomials.
These identities lead to a convenient formulation of the string equations.
As an application, we show that in the continuum limit the free energy of certain exactly solvable models
like the linear and double Penner models can be written as a sum of gaussian contributions
plus linear terms. To illustrate the one-cut case we discuss the linear, double and cubic Penner models,
and for the two-cut case we discuss theoretically and numerically the existence of a double-branch
structure of the free energy for the gaussian Penner model.
\end{abstract}
\pacs{05.90.+m}
\maketitle
%%%%%%%%%%%%%%%%%%%%%%%%%%%%%%%%%%%%%%%%%%%%%%%%%%%%%%%%%%%%%%%%%
%% INTRODUCTION %%%%%%%%%%%%%%%%%%%%%%%%%%%%%%%%%%%%%%%%%%%%%%%%%%%%%%
%%%%%%%%%%%%%%%%%%%%%%%%%%%%%%%%%%%%%%%%%%%%%%%%%%%%%%%%%%%%%%%%%
\section{Introduction}
In this paper we study the partition functions of random matrix models
\begin{equation}
	\label{mm}
	Z_{n,N}
	=
	\frac{1}{n!}
	\int_{\gamma\times\cdots\times\gamma}
	\prod_{j<k}(\lambda_j-\lambda_k)^2
	\exp\left(-N\sum_{i=1}^n W(\lambda_i)\right)
	\prod_{i=1}^n {\rm d}\lambda_i,
\end{equation}
with potentials $W(z)$ of Penner type
\begin{equation}
	\label{pe}
	W(z) = W_0(z)-\sum_{i=1}^k \mu_i \log (z-q_i),
\end{equation}
where $W_0(z)$ is a polynomial. We assume that $N>0$ and $\mu_i$ are real numbers, and 
that $q_i$ are arbitrary complex numbers. The choice of allowable integration contours $\gamma$
is nontrivial and must be discussed case by case. We also study the continuum limit
(often called large $N$ limit or {}'t~Hooft limit) of the corresponding free energy $F_{n,N}=\log Z_{n,N}$.
This continuum limit is defined by
\begin{equation}
	\label{eq:cl}
	n,N\rightarrow \infty,\quad x=\frac{n}{N}=\mbox{fixed},
\end{equation}
and underlies many of the applications of these models in theoretical
physics~\cite{PE88,DI90,DI91,AM94,DE02,DE03,MA06,CH07,DA10,BH13,PA95,MA94,DI99,SC14,EG10,SC10,SE44,KH93,KO99}.

The partition functions of several Penner models can be explicitly calculated using Selberg's
integral or some of its consequences~\cite{ME91}, and turn out to be products and quotients of
Euler's gamma functions. However,  the standard approach for studying the partition functions of
matrix models and their continuum limit  is the method of orthogonal polynomials ~\cite{BE80,DI95},
\begin{equation}
	\label{po}
	\int_{\gamma}P_{n,N}(z) P_{m,N}(z) \rme^{-N W(z)}  \rmd  z =\delta_{nm} h_{m,N},\quad n\geq 0,
\end{equation}
where $P_{n,N}(x) = x^n + \cdots$. For polynomial potentials in the one-cut case this method  has led
to rigorous proofs of the existence of the continuum limit expansion and to the determination
of its coefficients~\cite{BE80,DI95,BE79,BL05,AL11,ER08,ER09}.
Nevertheless, the application of the  method of orthogonal polynomials to Penner models~\cite{DI90,DI91,DE02,TA92}
is not so well established and even has been considered dubious (see, for example, the remarks
in appendix~2 of~\cite{SC10}).  The main difficulty lies in  the explicit formulation of the \emph{string equations} 
for determining the recurrence coefficients $r_{n,N}$ and $s_{n,N}$ in the three-term recursion relation 
\begin{equation}
	\label{rec} 
	z P_{n,N}(z) = P_{n+1,N}(z) + s_{n,N} P_{n,N}(z) + r_{n,N}  P_{n-1,N}(z).
\end{equation}
The string  equations should be expressed as difference equations for the recurrence coefficients and this is not easily done for 
Penner models due to the presence of  matrix elements  of the resolvent $(L-z)^{-1}$, where $L$ is the Jacobi matrix involved
in (\ref{rec}).  Thus, to the authors knowledge,  for Penner models there are not versions of methods to determine the form of
the  string equations  such as the \emph{summations of paths over a staircase} of Bessis, Itzykson and
Zuber~\cite{BE80,IT80,SH95,SH96}. The same situation arises with  other alternatives to the string equations
to compute  the continuum limit of the recurrence coefficients like the partial differential scheme of  Ercolani, McLaughlin and 
Pierce~\cite{ER08}. 

The present paper provides an implementation of the method of orthogonal polynomials  which  is particularly suitable
to deal with the string equations of Penner models, to obtain their solutions in the exactly solvable cases, and to determine
the coefficients of  their perturbative solutions in the continuum limit~(\ref{eq:cl}).

Our analysis extends previous work pertaining to matrix models with polynomial potentials~\cite{AL11,MA07,MM08,ALA11}.
The main idea is to combine the standard string equations with certain identities for the resolvent $(L-z)^{-1}$.
These  \emph{resolvent identities} are familiar in the theory of integrable systems~\cite{GE75i} (see also~\cite{KU85,JA88}
for the case of the Toda chain hierarchy), where they are
used to determine hierarchies of conserved densities. The idea is rather natural  because $L$ is the Lax operator
of the semi-infinite Toda chain hierarchy~\cite{DE99} (see  in particular~\cite{HA93} for the relationship between the
Toda hierarchy and some Penner matrix models).

Moreover, as it is well-known in the theory of integrable systems, the resolvent identities can be recurrently solved.
As a consequence we prove that they can be applied  to  compute  the partition functions of the exactly solvable Penner models,
so that they provide   an alternative derivation of the exact results obtained via the Selberg's
integral. Furthermore, we formulate the continuum limit of these resolvent identities and prove that they lead to a
 perturbative method to determine the continuum limit of non-exactly solvable models.

We organize our discussion as follows.
In section~\ref{sec:dse} we outline the standard method of orthogonal polynomials to determine partition functions
of matrix models. Then we derive a  combined system of string  equations and resolvent identities
for finding the recurrence coefficients  of orthogonal polynomials associated to Penner like potentials.
The reduction of the combined system to the case of $Z_2$-symmetric Penner models is also given. Section~\ref{sec:esp}
deals with three important examples of Penner models which exhibit  exactly solvable string equations: the linear,
gaussian and double Penner models. We illustrate in detail how the resolvent identities apply 
to solve the string equations  and how  the corresponding  partition functions  can be expressed 
in terms of the Gamma function and the Barnes $G$ function. In the case  of the double Penner model our calculation
of its partition function, which  represents a basic correlation function in conformal field theory~\cite{SC10},
provides a simple alternative to  other  derivations based on the Selberg integral~\cite{ME91} or on the tabulated
expressions of the normalization constants of Jacobi polynomials~\cite{SC10,AS72}. Section~\ref{sec:lnl} presents
a scheme to determine the continuum limit expansions of Penner models in the one-cut case.
We formulate the continuum limit of both the string equations and the resolvent identities  and provide a perturbative method
to derive the same type of expansions for the recurrence coefficients as those rigorously proved~\cite{BL05,KU00}
for polynomial potentials. For the sake of simplicity the technical details concerning existence and uniqueness  questions
are relegated to appendices~A and B. We take advantage of the exact expressions of the partition functions of the linear
and double Penner models  to confirm our results. At this point we notice that the free energy $\mathcal{F}$
of these models in the continuum limit is such that its second-order derivative $\partial_{xx}\mathcal{F}$ can be
decomposed as a sum of gaussian contributions $\partial_{xx}\mathcal{F}^G$. In Section~\ref{sec:lnlz2} we extend our
scheme to determine perturbative solutions of the string equations plus resolvent identities system in the large $N$
limit for $Z_2$-symmetric Penner models in the two-cut case. We assume a two-branch expansion for the recurrence
coefficient, show how it leads to a two-branch structure  of the large~$N$ expansion of the free energy, and illustrate
these results with a numerical computation of the corresponding exact values of the free energy.
Finally, in section~\ref{sec:cr} we briefly summarize the paper and point out possible extensions of this approach.
%%%%%%%%%%%%%%%%%%%%%%%%%%%%%%%%%%%%%%%%%%%%%%%%%%%%%%%%%%%%%%%%%
%% STRING EQUATIONS %%%%%%%%%%%%%%%%%%%%%%%%%%%%%%%%%%%%%%%%%%%%%%%%%%%
%%%%%%%%%%%%%%%%%%%%%%%%%%%%%%%%%%%%%%%%%%%%%%%%%%%%%%%%%%%%%%%%%
\section{Discrete string equations\label{sec:dse}}
In this section we first recall briefly the standard method of orthogonal polynomials~\cite{BE80,DI95},
and then discuss our implementation for Penner matrix models.

We recall that the partition function~(\ref{mm}) may be written as a product of the
normalization coefficients $h_{k,N}$,
\begin{equation}
	\label{mmc}
	Z_{n,N} = \prod_{k=0}^{n-1} h_{k,N},
\end{equation}
and that the ratios of successive normalization coefficients are the coefficients $r_{k,N}$ in the three-term
recursion relation~(\ref{rec}),
\begin{equation}
	\label{eq:rhh}
	r_{k,N} = \frac{h_{k,N}}{h_{k-1,N}}.
\end{equation}
For later reference we use~(\ref{eq:rhh}) to write the partition function $Z_{n,N}$ in terms of the first normalization constant
\begin{equation}
	\label{eq:h0}
	h_{0,N} = \int_\gamma \rme^{-N W(z)}\rmd z,
\end{equation}
and of the recurrence coefficients $r_{k,N}$:
\begin{equation}
	\label{eq:zhr}
	Z_{n,N} = h_{0,N}^n \prod_{k=1}^{n-1} r_{k,N}^{n-k}.
\end{equation}
Thus, the study of the partition function is reduced to the study of the recursion coefficients $r_{k,N}$.
For notational simplicity, hereafter the dependence of all the quantities
$Z_{n,N}$, $P_{n,N}$, $r_{n,N}$ and $s_{n,N}$ on $N$ will not be indicated explicitly.
%%%%%%%%%%%%%%%%%%%%%%%%%%%%%%%%%%%%%%%%%%%%%%%%%%%%%%%%%%%%%%%%%
\subsection{String equations and resolvent identities}
%%%%%%%%%%%%%%%%%%%%%%%%%%%%%%%%%%%%%%%%%%%%%%%%%%%%%%%%%%%%%%%%%
The three-term recursion relation (\ref{rec}) can be rewritten in matrix form as
\begin{equation}
	\label{rec1}
	L\left(\begin{array}{c} P_0(z) \\P_1(z) \\ \vdots \\P_k(z) \\ \vdots \end{array}\right)
	=
	z
	\left(\begin{array}{c} P_0(z) \\P_1(z) \\ \vdots \\P_k(z) \\ \vdots \end{array}\right),
\end{equation}
where $L$ is the Jacobi matrix~\cite{DE99}
\begin{equation}
	\label{jac}
	L = \left(\begin{array}{cccc} s_0 & 1 & 0 & \cdots \\
	                                            r_1 & s_1 & 1 & \cdots \\
	                                            0 & r_2 & s_2 & \cdots \\
	                                            \vdots & \vdots & \vdots & \ddots
	       \end{array}\right).
\end{equation}
In the method of orthogonal polynomials the recurrence coefficients $r_k$ and $s_k$ are
determined by the two \emph{string equations}
\begin{equation}
	\label{eq:str1}
	W'(L)_{n n-1} = \frac{n}{N},\quad n>1,
\end{equation}
\begin{equation}
	\label{eq:str2}
 	W'(L)_{n n} = 0,\quad n\geq 0.
\end{equation}
These equations are simple consequences of~(\ref{po}), (\ref{rec}) and of the assumption
\begin{equation}
	\int_{\gamma}\frac{ \rmd }{\rmd z} \left(P_j(z) P_k(z) \rme^{-N W(z)}\right)  \rmd  z = 0,
	\quad
	j,k\geq 0.
	\label{eq:id}
\end{equation}

At this point it becomes clear that the  resolvent operator $(L-z)^{-1}$ is a natural tool
in the theory of  Penner models, since we have to consider the elements $(n,n)$ and $(n,n-1)$
of the matrices
\begin{equation}
	W'(L) = W'_0(L) -\sum_{k= 1}^m \frac{\mu_k}{L-q_k}.
\end{equation}
Consequently we introduce the functions
\begin{equation}
	\label{genr}
	R_n(z) = \left(\frac{1}{z-L}\right)_{nn},
\end{equation}
\begin{equation}
	\label{gent}
	T_n(z)=1+2 \left(\frac{1}{z-L}\right)_{nn-1},
\end{equation}
and their respective expansions as $z\rightarrow\infty$
\begin{equation}
	\label{expr}
	R_n(z) = \frac{1}{z} \left(1+\sum_{k=1}^{\infty}\frac{(L^k)_{nn}}{z^k}\right),
\end{equation}
\begin{equation}
	\label{expt}
	T_n(z) = 1+\frac{2}{z} \sum_{k=1}^{\infty}\frac{(L^k)_{nn-1}}{z^k}.
\end{equation}
Thus, we can rewrite the string equations~(\ref{eq:str1})--(\ref{eq:str2}) as
\begin{equation}
	\label{str01}
	\frac{1}{2\pi\rmi}\oint_{\gamma_{\infty}}W_0'(z) T_n(z){\rmd z}+\sum_{i=1}^m \mu_i \left(T_n(q_i)-1\right)
	=
	\frac{2 n}{N}, \quad n\geq 1,
\end{equation}
\begin{equation}
	\label{str02}
	\frac{1}{2\pi\rmi}\oint_{\gamma_{\infty}}W_0'(z) R_n(z){\rmd z}+\sum_{i=1}^m \mu_i R_n(q_i)
	=
	0,\quad n\geq 0,
\end{equation}
where $\gamma_{\infty}$ is a large positively oriented circle around the origin.

Furthermore, in appendix~A we derive the resolvent identities
\begin{equation}
	\label{eq:rid1}
	T_n^2(z)-4 r_n R_n(z) R_{n-1}(z)=1,\quad n\geq 1,
\end{equation}
\begin{equation}
	\label{eq:rid2}
		2 (z-s_n) R_n(z)=T_{n+1}(z)+T_n(z),\quad n\geq 0,
\end{equation}
which allow us to compute $T_n(q_i)$,  $R_n(q_i)$, the expansions~(\ref{expr}) for $R_n(z)$ and~(\ref{expt})
for $T_n(z)$ in terms of the recurrence coefficients.
Therefore~(\ref{str01})--(\ref{eq:rid2}) constitute a system of equations for  the recurrence coefficients. These equations are
exactly solvable only in very special cases, some of which will be discussed in section~\ref{sec:esp}.
In sections~\ref{sec:lnl} and~\ref{sec:lnlz2} we will present a general perturbative scheme to solve them in the continuum limit.
%%%%%%%%%%%%%%%%%%%%%%%%%%%%%%%%%%%%%%%%%%%%%%%%%%%%%%%%%%%%%%%%%
\subsection{String equations for $Z_2$-symmetric Penner models}
%%%%%%%%%%%%%%%%%%%%%%%%%%%%%%%%%%%%%%%%%%%%%%%%%%%%%%%%%%%%%%%%%
For $Z_2$-symmetric Penner models with potentials of the form
\begin{equation}
	\label{pee}
	W(z) = W_0(z)-\sum_{i= 1}^m \mu_i  \log (z^2-q_i^2),\quad W_0(-z)=W_0(z),
\end{equation}
and defined over paths $\gamma$ symmetric under $z\rightarrow -z$, we have that $P_n(-z)=(-1)^n P_n(z)$.
In this case it follows from equations~(\ref{Rint}) and~(\ref{Tint}) of appendix~A that 
\begin{equation}
	\label{sim}
	T_n(-z)=T_n(z),\quad R_n(-z)=-R_n(z).
\end{equation}
As a consequence the second string equation~(\ref{str02}) is trivially satisfied,
while the first string equation~(\ref{str01}) takes the form
\begin{equation}
	\label{stre1}
	\frac{1}{2\pi\rmi}\oint_{\Gamma_{\infty}} W_0'(\lambda) T_n(\lambda){\rmd\lambda}
	+
	\sum_{i=1}^m \mu_i \left(T_n(q_i^2)-1\right)=\frac{n}{N},
\end{equation}
where $\lambda=z^2$, the function $T_n$ is expressed as a function of $\lambda$, and $\Gamma_{\infty}$ is a large
positively oriented circle around the origin in the $\lambda$ plane.  Moreover,  taking into account that $s_n=0$,
the relations~(\ref{eq:rid1})--(\ref{eq:rid2}) reduce to
\begin{equation}
	\label{eq:rid1n}
	T_n^2(z)-4 r_n R_n(z) R_{n-1}(z)=1,\quad n\geq 1,
\end{equation}
\begin{equation}
	\label{eq:rid2n}
		2 z \,R_n(z)=T_{n+1}(z)+T_n(z),\quad n\geq 0,
\end{equation}
which obviously imply the following equation involving only $T_n(z)$ and $r_n$,
\begin{equation}
	\label{recce1}
	z^2 \left(T_n^2(z)-1\right) = r_n\left(T_{n+1}(z)+T_n(z)\right)\left(T_{n-1}(z)+T_n(z)\right).
\end{equation}
%%%%%%%%%%%%%%%%%%%%%%%%%%%%%%%%%%%%%%%%%%%%%%%%%%%%%%%%%%%%%%%%%
%% PENNER %%%%%%%%%%%%%%%%%%%%%%%%%%%%%%%%%%%%%%%%%%%%%%%%%%%%%%%%%%
%%%%%%%%%%%%%%%%%%%%%%%%%%%%%%%%%%%%%%%%%%%%%%%%%%%%%%%%%%%%%%%%%
\section{Exactly solvable Penner models\label{sec:esp}}
%%%%%%%%%%%%%%%%%%%%%%%%%%%%%%%%%%%%%%%%%%%%%%%%%%%%%%%%%%%%%%%%%
In this section we show how equations~(\ref{str01})--(\ref{eq:rid2}) allow us to rederive in a simple form
the explicit expressions for the partition functions of some exactly solvable models.
%%%%%%%%%%%%%%%%%%%%%%%%%%%%%%%%%%%%%%%%%%%%%%%%%%%%%%%%%%%%%%%%%
\subsection{The Barnes $G$ function}
%%%%%%%%%%%%%%%%%%%%%%%%%%%%%%%%%%%%%%%%%%%%%%%%%%%%%%%%%%%%%%%%%
As we anticipated in the Introduction, the partition functions of several exactly solvable Penner models are products
and quotients of Euler gamma functions. These expressions can be written more concisely in terms of the
Barnes $G$ function~\cite{BA00,OL10}, which satisfies
\begin{equation}
	\label{eq:gzp1}
	G(z+1)=\Gamma(z) G(z),\quad G(1)=1.
\end{equation}
From these equations it follows immediately that
\begin{equation}
	\label{id2}
	\prod_{k=0}^{n-1}\Gamma(j k+\alpha+1)\Gamma(j k+\alpha+2)\cdots\Gamma(j k+\alpha+j)
	=
	\frac{G(j n+\alpha+1)}{G(\alpha+1)},
\end{equation}
which in the particular case $j=1$ implies 
\begin{equation}
	\label{id1}
	\prod_{k=1}^{n-1}(k+\alpha)^{n-k} = \frac{G(n+\alpha+1)}{G(\alpha+1) \Gamma(\alpha+1)^n}.
\end{equation}

In our analysis of the continuum limit we will also need the asymptotic expansion for large $z$ of the Barnes $G$
function. This expansion is often written keeping an unexpanded gamma function
(see equation~5.17.5 in~\cite{OL10}), but we find more convenient the fully expanded version
\begin{eqnarray}
	\label{aba}
	\log G(z+1) & \approx &\frac{1}{2}z^2 \log z-\frac{3}{4}z^2+\frac{1}{2}z\log(2\pi)-\frac{1}{12}\log z+\zeta'(-1)
	\nonumber\\
			  &     &  {}+ \sum_{k= 2}^\infty\frac{B_{2k}}{2k(2k-2)}\frac{1}{z^{2k-2}}, \quad z\rightarrow\infty,
\end{eqnarray}
where $B_{2k}$ are the Bernoulli numbers and $\zeta'$ is the derivative of the Riemann zeta function.
Incidentally, the $\chi_k=B_{2k}/(2k(2k-2))$ which appear in~(\ref{aba}) are precisely
the virtual Euler characteristic numbers for the moduli space of nonpunctured Riemann surfaces~\cite{HA86}.
This property is at the root of the appearance of several exactly solvable matrix models in
noncritical $c=1$ string theory~\cite{PA10}.
%%%%%%%%%%%%%%%%%%%%%%%%%%%%%%%%%%%%%%%%%%%%%%%%%%%%%%%%%%%%
\subsection{The gaussian model}
%%%%%%%%%%%%%%%%%%%%%%%%%%%%%%%%%%%%%%%%%%%%%%%%%%%%%%%%%%%%
The simplest exactly solvable matrix model is the gaussian model $W(z)=z^2/2$, which, in a certain
sense discussed in section~\ref{sec:lnl}, can be considered as a building block of several Penner models.

The recurrence coefficient for the gaussian model is $r_k=k/N$. Equation~(\ref{eq:h0}) gives the lowest
normalization constant
\begin{equation}
	h_0 = \int_{-\infty}^\infty \rme^{-N z^2/2}\rmd z = \left(\frac{2\pi}{N}\right)^{1/2},
\end{equation}
and~(\ref{eq:zhr}) and~(\ref{id1}) with $\alpha=0$ lead to 
\begin{equation}
	Z_n = \left(\frac{2\pi}{N}\right)^{n/2} \prod_{k=1}^{n-1} \left(\frac{k}{N}\right)^{n-k}
	       = \frac{(2\pi)^{n/2}}{N^{n^2/2}}  G(n+1).
\end{equation}
Therefore, for fixed $x=n/N$ and $\epsilon=1/N\to 0$, equation~(\ref{aba}) implies the asymptotic expansion
\begin{eqnarray}
	\label{gau}
	F^G(\epsilon,x) & \approx & \frac{1}{\epsilon^2}\frac{x^2}{2}\left(\log x-\frac{3}{2}\right)
	                 + \frac{1}{\epsilon}x \log (2\pi)
	                 + \frac{1}{12}\log \epsilon
	                 + \zeta'(-1)-\frac{1}{12}\log x \nonumber \\
                &    & {}+ \sum_{k=1}^\infty\epsilon^{2k}\frac{B_{2k+2}}{4k(k+1)}\frac{1}{x^{2k}}.
\end{eqnarray}
%%%%%%%%%%%%%%%%%%%%%%%%%%%%%%%%%%%%%%%%%%%%%%%%%%%%%%%%%%%%
\subsection{The linear Penner model}
%%%%%%%%%%%%%%%%%%%%%%%%%%%%%%%%%%%%%%%%%%%%%%%%%%%%%%%%%%%%
The linear Penner model is defined by the potential
\begin{equation}
	W(z) = z-\log z,
\end{equation}
and the integration contour $\gamma$ is the nonnegative real axis. It was first introduced to study the orbifold
Euler characteristic of the moduli space of $n$-punctured Riemann surfaces at genus $g$~\cite{PE88}
and it is used in $c=1$ noncritical string theory~\cite{DI90,DI91}. 

Since $N>0$, the weight $\rme^{-N W(x)}$ vanishes both at $x=0$ and at $x=\infty$, condition~(\ref{eq:id})
is satisfied, and the corresponding string equations~(\ref{str01})--(\ref{str02}) read
\begin{equation}
	T_n(0)-1=\frac{2 n}{N},
	\label{rsp0}
\end{equation}
\begin{equation}
	1+R_n(0)=0.
	\label{rsp1}
\end{equation}
Therefore $T_n(0)=1+2 n/N$ and $R_n(0)=-1$. Furthermore, the resolvent
identities~(\ref{eq:rid1})--(\ref{eq:rid2}) at $z=0$ are
\begin{equation}
	\label{reccc1}
	T_n^2(0)-4 r_n R_n(0) R_{n-1}(0)=1,
\end{equation}
\begin{equation}
	\label{reccc2}
	-2 s_n R_n(0)=T_{n+1}(0)+T_n(0),
\end{equation}
and we obtain immediately
\begin{equation}
	\label{rp}
	r_n=\frac{n}{N}+\frac{n^2}{N^2},
\end{equation}
\begin{equation}
	\label{sp}
	s_n=1+\frac{2 n+1}{N}.
\end{equation}
The lowest normalization constant is
\begin{equation}
	h_0 = \int_0^{\infty} \rme^{-N(x-\log x)}\rmd x
	       = \frac{\Gamma(N+1)}{N^{N+1}},
\end{equation}
and~(\ref{eq:zhr}) and~(\ref{id1}) lead to 
\begin{equation}
	\fl
	\label{pz}
	Z_n = \left( \frac{\Gamma(N+1)}{N^{N+1}}\right)^n
	           \prod_{k=1}^{n-1}
	           \left(\frac{k}{N}+\frac{k^2}{N^2}\right)^{n-k}
              =  \frac{1}{N^{n(N+n)}}\frac{G(n+1)G(N+n+1)}{G(N+1)}.
\end{equation}
%%%%%%%%%%%%%%%%%%%%%%%%%%%%%%%%%%%%%%%%%%%%%%%%%%%%%%%%%%%%
\subsection{The $Z_2$-symmetric gaussian Penner model}
%%%%%%%%%%%%%%%%%%%%%%%%%%%%%%%%%%%%%%%%%%%%%%%%%%%%%%%%%%%%
The symmetric gaussian Penner model is defined by the potential
\begin{equation}
	W(z )= \frac{z^2}{2}-\frac{1}{2}\log z^2,
\end{equation}
where the integration contour $\gamma$ is the whole real axis.
This model has been applied to derive new asymptotics for the generalized Laguerre polynomials~\cite{DE03},
to the study of structural glasses in the high temperature phase~\cite{DE02}
and to the study of interacting RNA folding and structure combinatorics~\cite{BH13}.

Since $N>0$, along the real axis the weight $\rme^{-N W(z)}$ vanishes at $x=\pm\infty$,
and the condition~(\ref{eq:id}) is satisfied. In this case the string equation~(\ref{stre1}) reads
\begin{equation}
	\label{rspe}
	2 r_n+T_n(0)-1 = \frac{2 n}{N}.
\end{equation}
Taking into account that~(\ref{sim}) implies $R_n(0)=0$, the resolvent identities~(\ref{eq:rid1n})--(\ref{eq:rid2n})
at $z=0$ become
\begin{equation}
	\label{recceg}
	T_n^2(0)=1,\quad T_{n+1}(0)+T_n(0)=0.
\end{equation}
Then, using the boundary condition $r_0=0$, we deduce that
\begin{equation}
	T_n(0) = (-1)^n,
\end{equation}
and
\begin{equation}
 \label{rgs}
	r_n = \frac{n}{N}+\frac{1}{2}\left(1-(-1)^n\right).
\end{equation}
Proceeding as in the previous cases, we find that the lowest normalization constant is
\begin{equation}
	\label{h0gp}
	h_0=\int_{-\infty}^\infty\rme^{-N(z^2-\log(z^2))/2}\rmd z
	      =\left(\frac{2}{N}\right)^{\frac{N+1}{2}}  \Gamma\left(\frac{N+1}{2}\right).
\end{equation}
However, due to the term $(-1)^n$ in the expression~(\ref{rgs}) for the coefficient $r_n$, the
simplest way to calculate the partition function is to treat separately the even and the odd terms.
Each case can be handled straightforwardly with the following results,
\begin{equation}
	\label{gapo}
	Z_{2k+1} = \left(\frac{2}{N}\right)^{(2k+1)(k+(N+1)/2)}
	                   \frac{k! G(k+1)^2  G\left(\frac{N+1}{2}+k+1\right)^2  \Gamma \left(\frac{N+1}{2}\right)}{
	                           G\left(\frac{N+1}{2}+1\right)^2},
\end{equation}
and
\begin{equation}
	\label{gape}
	Z_{2k} = \left(\frac{2}{N}\right)^{k(2k+N)}
	               \frac{G(k+1)^2  G\left(\frac{N+1}{2}+k+1\right)^2\Gamma \left(\frac{N+1}{2}\right)}{
	                       G\left(\frac{N+1}{2}+1\right)^2  \Gamma\left(\frac{N+1}{2}+k\right)},
\end{equation}
which using repeatedly equation~(\ref{eq:gzp1}) for $G(z+1)$ can be merged back into the single expression
\begin{eqnarray}
	\label{dpz}
	\fl
	Z_{n} & = & \left(\frac{2}{N}\right)^{n(n+N)/2}\times\nonumber\\
	\fl
	          &    & \frac{  G\left(\frac{2n+3+(-1)^n}{4}\right)
	                             G\left(\frac{2n+5-(-1)^n}{4}\right)
	                             G\left(\frac{2n+3-(-1)^n+2N}{4}\right)
	                             G\left(\frac{2n+5+(-1)^n+2N}{4}\right)}{
	                             G\left(\frac{N+1}{2}\right)
	                             G\left(\frac{N+3}{2}\right)}.
\end{eqnarray}
%%%%%%%%%%%%%%%%%%%%%%%%%%%%%%%%%%%%%%%%%%%%%%%%%%%%%%%%%%%%%%
\subsection{The double Penner model}
%%%%%%%%%%%%%%%%%%%%%%%%%%%%%%%%%%%%%%%%%%%%%%%%%%%%%%%%%%%%%%
The double Penner model is defined by the potential
\begin{equation}
        \label{dps}
	W(z) = -\mu_0 \log z-\mu_1\log(1-z),
\end{equation}
where the principal branches of the logarithmic function are understood (i.e., $-\pi<\im(\log z)<\pi$),
and where the integration contour $\gamma$ is the $[0,1]$ interval of the real axis.
It has been used  to study  the  critical behavior of the Kazakov-Migdal model in $U(N)$ Lattice Gauge
theory~\cite{PA95} and to describe its tricritical point associated to a Kosterlitz-Thouless phase 
transitions~\cite{MA94}.  Incidentally, multi-Penner  matrix models of the form 
\begin{equation}
	\label{mup}
 	W(z)= -\sum_{i=1}^k \mu_i \log(z-q_i),
\end{equation}
are an active area of research because of the remarkable connections between matrix models,
conformal field theory and supersymmetric gauge theories~\cite{DI99,SC14,EG10,SC10,SE44,KH93,KO99}.
In particular, correlation functions in these field theories turn to be  described  by partition functions
$Z_{n,N}$ of  models with potentials of the form~(\ref{mup}). 

Note that along the interval $[0,1]$ we have $\rme^{-N W(z)} = x^{\mu_0 N}(1-x)^{\mu_1 N}$.
Therefore, if $\mu_0 N>0$ and $\mu_1 N>0$ the weight vanishes at $x=0$ and at $x=1$, and the
identity~(\ref{eq:id}) is valid. The corresponding string equations~(\ref{str01})--(\ref{str02}) read
\begin{equation}
	\label{rsdp1}
	\mu_0 (T_n(0)-1)+\mu_1 (T_n(1)-1)=\frac{2 n}{N},
\end{equation}
\begin{equation}
	\label{rsdp2}
	\mu_0 R_n(0)+\mu_1 R_0(1)=0,
\end{equation}
and the resolvent identities~(\ref{eq:rid1})--(\ref{eq:rid2}) at $z=0$ and at $z=1$ lead to 
\begin{equation}
	\label{recc1}
	T_n^2(0)-4 r_n R_n(0) R_{n-1}(0)=1,
\end{equation}
\begin{equation}
	\label{recc2}
	-2 s_n R_n(0)=T_{n+1}(0)+T_n(0),
\end{equation}
\begin{equation}
	\label{recc3}
	T_n^2(1)-4 r_n R_n(1) R_{n-1}(1)=1,
\end{equation}
\begin{equation}
	\label{recc4}
	2 (1-s_n) R_n(1)=T_{n+1}(1)+T_n(1).
\end{equation}
This system~(\ref{rsdp1})--(\ref{recc4}) is exactly solvable. In terms of $\alpha_0=\mu_0 N$ and $\alpha_1=\mu_1 N$,
the solution is given by
\begin{equation}
	R_n(0) = -(2 n+1+\alpha _0+\alpha _1)/\alpha_0,
\end{equation}
\begin{equation}
        R_n(1) = -(2 n+1+\alpha _0+\alpha _1)/\alpha_1,
\end{equation}
 \begin{equation}
       T_n(0) = \frac{2n^2+(2n+\alpha_0)(\alpha_0+\alpha_1)}{\alpha_0(2n+\alpha_0+\alpha_1)},
\end{equation}
\begin{equation}
       T_n(1) = \frac{2n^2+(2n+\alpha_1)(\alpha_0+\alpha_1)}{\alpha_1(2n+\alpha_0+\alpha_1)},
\end{equation}
and
\begin{equation}
	r_n = \frac{n(n+\alpha_0)(n+\alpha_1)(n+\alpha_0+\alpha_1)}{
	                 (2n+\alpha_0+\alpha_1)^2(2n+\alpha_0+\alpha_1-1)(2n+\alpha_0+\alpha_1+1)},
	\label{eq:dprn}
\end{equation}
\begin{equation}
	s_n = \frac{2n^2+2n(\alpha_0+\alpha_1+1)+(\alpha_0+\alpha_1)(\alpha_0+1)}{
	                  (2n+\alpha_0+\alpha_1)(2n+\alpha_0+\alpha_1+2)}.
	                  \label{eq:dpsn}
\end{equation}
From the expression~(\ref{eq:dprn}), using the Legendre duplication formula 
\begin{equation}
	\Gamma(z) \Gamma(z+1/2) = 2^{1-2 z} \sqrt{\pi} \Gamma(2 z),
\end{equation}
and the zero-th normalization coefficient
\begin{equation}
	h_0 = \int_0^1 x^{\alpha_0}(1-x)^{\alpha_1}\rmd x
	=
	\frac{\Gamma(\alpha_0+1)\Gamma(\alpha_1+1)}{\Gamma(\alpha_0+\alpha_1+2)},
\end{equation}
it follows immediately that the partition function takes the form
\begin{equation}
	\label{doup}
	Z_n = \frac{G(n+1)G(n+\alpha_0+1)G(n+\alpha_1+1)G(n+\alpha_0+\alpha_1+1)}{
	                         G(\alpha_0+1)G(\alpha_1+1)G(2n+\alpha_0+\alpha_1+1)}.
\end{equation}
%%%%%%%%%%%%%%%%%%%%%%%%%%%%%%%%%%%%%%%%%%%%%%%%%%%%%%%%%%%%%%%%%
%% LARGE N %%%%%%%%%%%%%%%%%%%%%%%%%%%%%%%%%%%%%%%%%%%%%%%%%%%%%%%%%
%%%%%%%%%%%%%%%%%%%%%%%%%%%%%%%%%%%%%%%%%%%%%%%%%%%%%%%%%%%%%%%%%
\section{The continuum limit in the one-cut case\label{sec:lnl}}
%%%%%%%%%%%%%%%%%%%%%%%%%%%%%%%%%%%%%%%%%%%%%%%%%%%%%%%%%%%%%%%%%
The method of orthogonal polynomials provides a general perturbative scheme to generate the expansion of the partition
function of matrix models with polynomial potentials in the continuum limit~\cite{DI95,BE79,BL05}.
In this section we apply this scheme to Penner matrix models in the one-cut case.

Equations~(\ref{mmc}) and~(\ref{eq:rhh}) imply that
\begin{equation}
	\label{main}
	\frac{Z_{n+1}  Z_{n-1}}{Z_{n}^2} = r_n.
\end{equation}
Therefore, the continuum limit expansion of the free energy $F_n=\log Z_n$ can be related to the continuum
limit expansion of $r_n$. In appendix~B we show that in the continuum limit~(\ref{eq:cl}) the recurrence coefficients
$r_n$ and $s_n$ for Penner models in the one-cut case admit asymptotic expansions of the form
\begin{equation}
	\label{exr}
	r_{n} \sim r(\epsilon,x) \approx \sum_{k=0}^{\infty} \epsilon^{2 k} {\rho}_k (x),
\end{equation}
\begin{equation}
	\label{exs}
	s_{n} \sim s(\epsilon,x) \approx \sum_{k=0}^{\infty} \epsilon^{k} {\sigma}_k (x),
\end{equation}
where
\begin{equation}
	\epsilon = \frac{1}{N}\rightarrow 0,
	\quad
	x=\frac{n}{N}\mbox{ fixed.}
\end{equation}
Note that the expansion $r(\epsilon,x)$ contains only even powers of $\epsilon$, while
the expansion for $s(\epsilon,x)$ contains all powers of $\epsilon$ (see appendix~B).

Taking the logarithm of~(\ref{main}) and as a consequence of~(\ref{exr}) we deduce that
the continuum limit of the free energy  is represented  by an expansion  
\begin{equation}\label{freee}
F_n \sim F(\epsilon,x)
\end{equation}
which satisfies 
\begin{equation}
	F(\epsilon,x+\epsilon) + F(\epsilon,x-\epsilon) - 2 F(\epsilon,x)
	=
	\log r(\epsilon,x),
\end{equation}
or
\begin{equation}
	\label{e1}
	(\rme^{\frac{\epsilon}{2}\partial_x}-\rme^{-\frac{\epsilon}{2}\partial_x})^2 F(\epsilon,x)
	=
	4 \sinh^2\left(\frac{\epsilon}{2} \partial_x\right) F(\epsilon,x)=\log r(\epsilon,x).
\end{equation}
Therefore, using the identity
\begin{equation}
	\label{idb}
	\left(\frac{t}{2}\right)^2{\rm csch}^2\frac{t}{2}
	=
	1-\sum_{k=1}^{\infty} \frac{B_{2k}}{(2k) (2k-2)!} t^{2k},
\end{equation}
we find that $F(\epsilon,x)$  obeys  the differential equation 
\begin{equation}
	\label{main2}
	\epsilon^2 \partial_{xx} F(\epsilon,x)
	=
	\mathrm{L}(\epsilon\partial_x) \log r(\epsilon,x),
\end{equation}
where $\mathrm{L}(\epsilon\partial_x)$ denotes the vertex operator (infinite-order differential operator)
\begin{equation}
	\label{ope}
	\mathrm{L}(\epsilon \partial_x)
	=
	1-\sum_{k=1}^{\infty} \frac{B_{2k}}{(2k) (2k-2)!} (\epsilon \partial_x)^{2k}.
\end{equation}
Hence, it follows from~(\ref{main2}) that $F(\epsilon,x)$ can be written  as
\begin{equation}
	\label{expf}
	F(\epsilon,x)
	=
	F_{{\rm lin}}
	+
	\mathcal{F}(\epsilon,x),
\end{equation}
where $F_{{\rm lin}}$ is  an $\epsilon$-dependent term linear in $x$ and
\begin{equation}
	\label{expfcal}
	\mathcal{F}(\epsilon,x)
	=
	\frac{1}{\epsilon^2}\int_0^x (x-t)\log r(\epsilon,t) \rmd t
	-
	\sum_{k=1}^{\infty} \frac{ B_{2k}(\epsilon \partial_x)^{2k-2}}{(2k) (2k-2)!} \log r(\epsilon,x).
\end{equation}

Inserting the expansion~(\ref{exr}) of 
$r(\epsilon,x)$ into~(\ref{expf}) we obtain that $\mathcal{F}(\epsilon,x)$ has an asymptotic expansion in powers of
$\epsilon^2$
\begin{equation}
	\label{expf2}
	\mathcal{F}(\epsilon,x)
	\approx  \sum_{k=0}^\infty \epsilon^{2k-2}  {\mathcal F}_k(x),
\end{equation}
which is known as the  \emph{topological or genus expansion} of the free energy.
The coefficients ${\mathcal F}_k(x)$  can be expressed in terms of those of $r(\epsilon,x)$.  For example,
\begin{eqnarray}
	\label{efes0}
	 \fl
        {\mathcal F}_0 &=& \int_0^x (x-t) \log {\rho}_0(t)  \rmd  t,\\
        \fl
        \label{efes1}
        {\mathcal F}_1 &=& \int_0^x (x-t) \frac{{\rho}_1(t)}{{\rho}_0(t)} \rmd  t-\frac{1}{12}\log {\rho}_0(x),\\
        \fl
        {\mathcal F}_2 &= &\int_0^x (x-t)
                  \left(\frac{{\rho}_2(t)}{{\rho}_0(t)}-\frac{1}{2}\frac{{\rho}_1(t)^2}{{\rho}_0(t)^2}\right) \rmd  t
                  -
                  \frac{1}{12}\frac{{\rho}_1(x)}{{\rho}_0(x)} + \frac{1}{240}\partial_{xx}\log {\rho}_0(x).
\end{eqnarray}
As we will see below, for exactly solvable matrix models we can obtain an asymptotic expansion
of $F_{{\rm lin}}$ from the asymptotic expansions of the gamma and Barnes functions. 
In the general case one can derive the asymptotic expansion for $F_{{\rm lin}}$ by applying
the Euler-MacLaurin summation formula to  $F_n-F_n^G$ where $F_n^G$ is the gaussian
free energy.
%%%%%%%%%%%%%%%%%%%%%%%%%%%%%%%%%%%%%%%%%%%%%%%%%%%%%%%%%
\subsection{Decomposition into gaussian contributions}
%%%%%%%%%%%%%%%%%%%%%%%%%%%%%%%%%%%%%%%%%%%%%%%%%%%%%%%%%
It is worth noticing that the expression~(\ref{expfcal}) for $\mathcal{F}(\epsilon,x)$ is a linear functional of
$\log r(\epsilon,x)$. Therefore, whenever $r(\epsilon,x)$ is a rational function of $x$, the corresponding
expression of $\mathcal{F}(\epsilon,x)$ can be written as a sum of gaussian contributions
$\mathcal{F}^G(\epsilon a,a x + b)$ plus elementary integrals linear in $x$.
We  will see below that this property is satisfied by the linear and double Penner models. 

More concretely, the asymptotic expansion of the gaussian contribution $\mathcal{F}^G(\epsilon,x)$
corresponding to $r(\epsilon,x)=x$ in~(\ref{expfcal}), is
\begin{equation}
	\label{mcfg}
	\mathcal{F}^G(\epsilon,x)
	 \approx 
	\frac{1}{\epsilon^2} x^2
	\left(\frac{1}{2}\log x-\frac{3}{4}\right)
	- \frac{1}{12}\log x
	+\sum_{k=1}^\infty\frac{B_{2k+2}\epsilon^{2k}}{2k(2k+2)}\frac{1}{x^{2k}},	
\end{equation}
and the difference between this $\mathcal{F}^G$ and the expression of $F^G$ given
in~(\ref{gau}) are precisely the terms linear in $x$
\begin{equation}
	F^G_{\rm lin} = F^G - \mathcal{F}^G = \frac{1}{\epsilon} x \log(2\pi)
	+ \frac{1}{12}\log\epsilon +\zeta'(-1),
\end{equation}
whereas the function $\mathcal{F}(\epsilon,x)$  corresponding to $r(\epsilon,x)= a x + b$ is
\begin{equation}
	\label{axpb}
	\mathcal{F}(\epsilon,x)
	=
	\mathcal{F}^G(\epsilon a, a x+b)
	+
	\frac{b}{4 \epsilon^2 a^2} (3 b+4 a x-2 (b+2 a x) \log b).
\end{equation}
%%%%%%%%%%%%%%%%%%%%%%%%%%%%%%%%%%%%%%%%%%%%%%%%%%%%%%%%%%%%%%%%%
\subsection{Gaussian behavior}
%%%%%%%%%%%%%%%%%%%%%%%%%%%%%%%%%%%%%%%%%%%%%%%%%%%%%%%%%%%%%%%%%
We have just seen that for exactly solvable matrix models in which   $r(\epsilon,x)$ is a rational function of $x$, the free energy
is, up to linear terms, a sum of gaussian contributions $\mathcal{F}^G(\epsilon a,a x + b)$. Furthermore, since
\begin{equation}
	\label{axpb2}
\mathcal{F}^G(a \epsilon , a x+b)=\mathcal{F}^G\left(\epsilon ,  x+\frac{b}{a}\right)
	+\frac{\log a}{2 \epsilon^2} \left(x+\frac{b}{a}\right)^2
	-\frac{\log a}{12},
\end{equation}
the second-order derivative $\partial_{xx}\mathcal{F}(\epsilon,x)$ is, up to a constant, a linear combination
of the second-order derivatives of the gaussian contributions $\partial_{xx}\mathcal{F}^G(\epsilon , x +c_i)$. 
These constants $c_i$ are rather special since for $x$ near  each of the values $-c_i$ the behavior of the function
$\partial_{xx}\mathcal{F}(\epsilon,x)$ is dominated by a single gaussian term $\partial_{xx}\mathcal{F}^G(\epsilon , x +c_i)$.
%%%%%%%%%%%%%%%%%%%%%%%%%%%%%%%%%%%%%%%%%%%%%%%%%%%%%%%%%%%%%%%%%
\subsection{Solutions of the string equations in the continuum limit}
%%%%%%%%%%%%%%%%%%%%%%%%%%%%%%%%%%%%%%%%%%%%%%%%%%%%%%%%%%%%%%%%%
We next discuss our method to determine the coefficients $\rho_k(x)$ and $\sigma_k(x)$ of the
expansions~(\ref{exr})--(\ref{exs}) from the continuum limit of the string equations.

We prove in appendix~B that the continuum limit of the generating functions $R_n(z)$ and $T_n(z)$ is represented
by asymptotic expansions 
\begin{equation}
	\label{exR}
	R_n(z) \sim R(\epsilon,z,x) \approx \sum_{k=0}^{\infty} \epsilon^k R_k (z,x),
\end{equation}
\begin{equation}
	\label{exT}
	T_n(z) \sim T(\epsilon,z,x) \approx \sum_{k=0}^{\infty} \epsilon^{2k} T_k (z,x).
\end{equation}
Note that the expansion $R(\epsilon,z,x)$ contains all powers of $\epsilon$, while
the expansion for $T(\epsilon,z,x)$ contains only even powers of $\epsilon$ (see appendix~B).

The coefficients $R_k(z,x)$ and $T_k(z,x)$ can be recursively determined from the continuum limit of the recurrence
relations~(\ref{eq:rid1})--(\ref{eq:rid2}) which become
\begin{equation}
	\label{reccl1}
	T(\epsilon,z,x)^2-4 r(\epsilon,x)R(\epsilon,z,x) R(\epsilon,z,x-\epsilon)=1,
\end{equation}
and
\begin{equation}
	\label{reccl2}	
	                                 2 (z-s(\epsilon,x)) R(\epsilon,z,x)=T(\epsilon,z,x+\epsilon)+T(\epsilon,z,x).
\end{equation}
Therefore the coefficients $R_k(z,x)$ and $T_k(z,x)$ can be obtained in terms of $\rho_k$ and $\sigma_k$.
For  example,
\begin{equation}
	\label{R0}
	R_0(z,x) = w(z,x),
\end{equation}
\begin{equation}
	\label{T0}
	T_0(z,x)= (z-{\sigma}_0) w(z,x),
\end{equation}
where
\begin{equation}
	w(z,x) = \frac{1}{\sqrt{(z-{\sigma}_0(x))^2-4{\rho}_0(x)}}.
\end{equation}

In the continuum limit  the  string equations~(\ref{str01})--(\ref{str02}) are given by
\begin{equation}
	\label{str1}
		\frac{1}{2\pi\rmi}\oint_{\gamma_{\infty}}W_0'(z) T(\epsilon,z,x)\rmd z
		+
		\sum_{i=1}^m \mu_i \left(T(\epsilon,q_i,x)-1\right)=2 x,
\end{equation}		
\begin{equation}
	\label{str2}		
				\frac{1}{2\pi\rmi}\oint_{\gamma_{\infty}}W_0'(z) R(\epsilon,z,x)\rmd z
		+
		\sum_{i=1}^m \mu_i R(\epsilon,q_i,x)=0.
\end{equation} 
Inserting the expansions~(\ref{exR}) and~(\ref{exT}) into~(\ref{str1}) and~(\ref{str2})
and identifying coefficients of powers $\epsilon^{2k}$ leads the equations 
\begin{equation}
	\label{str1k}
			\frac{1}{2\pi\rmi}\oint_{\gamma_{\infty}} W_0'(z)T_k(z,x) \rmd z
		+
		\sum_{i=1}^m \mu_i (T_k(q_i,x)-\delta_{k0} )= 2\,\delta_{k0}\,x,
\end{equation}
\begin{equation}
\label{str12k}
		\frac{1}{2\pi\rmi}\oint_{\gamma_{\infty}} W_0'(z)R_{2k}(z,x) \rmd z
		+
		\sum_{i=1}^m \mu_i R_{2k}(q_i,x) = 0,
\end{equation}
which in turn determine recursively the coefficients ${\rho}_k (x)$ and ${\sigma}_k (x)$.
Furthermore, since $s(\epsilon,x)=s(-\epsilon, x+\epsilon)$ the coefficients $\sigma_k$ for odd $k$
are determined by the coefficients $\sigma_l$ for even $l$. Therefore for each  $k$ the
system (\ref{str1k})--(\ref{str12k}) determines the coefficients $\rho_k$ and $\sigma_{2k}$
in terms of lower order coefficients and their $x$ derivatives (see appendix~B). 

For example, taking into account~(\ref{R0})--(\ref{T0}) and setting $k=0$
in the system~(\ref{str1k})--(\ref{str12k}), we get a pair of equations  for the leading
coefficients $\rho_0$ and $\sigma_0$:
\begin{equation}
	\label{sysrs0}
	\fl
		\frac{1}{2\pi\rmi}\oint_{\gamma_{\infty}}
		(z-{\sigma}_0)w(z,x)W_0'(z)\rmd z
		+
		\sum_{i=1}^m \mu_i \left((q_i-{\sigma}_0)w(q_i,x)-1\right)
		=
		2x,
\end{equation}
\begin{equation}
	\label{sysrs02}	
	\fl
		\frac{1}{2\pi\rmi}\oint_{\gamma_{\infty}}
		w(z,x)W_0'(z)\rmd z
		+
		\sum_{i=1}^m \mu_i w(q_i,x)
		=
		0.
\end{equation}
These equations determine the so-called \emph{planar limit}  or \emph{zero genus contribution}
of the continuum expansion. In particular they can be used to calculate the endpoints of the
eigenvalue support $[a,b]$ since they are related to $\rho_0$ and $\sigma_0$ according to~\cite{TA92}
\begin{equation}
	a=\sigma_0-2\sqrt{\rho_0},\quad b=\sigma_0+2\sqrt{\rho_0}.
\end{equation}
%%%%%%%%%%%%%%%%%%%%%%%%%%%%%%%%%%%%%%%%%%%%%%%%%%%%%%%%%%%%%%%%%
\subsection{$Z_2$-symmetric Penner models in the one-cut case}
%%%%%%%%%%%%%%%%%%%%%%%%%%%%%%%%%%%%%%%%%%%%%%%%%%%%%%%%%%%%%%%%%
In the one-cut case, the continuum limit of the string equation~(\ref{stre1}) for $Z_2$-symmetric Penner models takes the form
\begin{equation}
	\label{stre11}
	\frac{1}{2\pi\rmi}\oint_{\Gamma_{\infty}}W_0'(\lambda) T(\epsilon,\lambda,x)\rmd\lambda
	+
	\sum_{i=1}^m \mu_i T(\epsilon,q_i^2,x)
	=
	x+\sum_{i=1}^m \mu_i,
\end{equation}
where $\lambda=z^2$ and $T$ is assumed to be a function of $\lambda$. Moreover, the relations~(\ref{reccl1})--(\ref{reccl2}) 
reduce to 
\begin{equation}
	\label{reccle1}
	\fl
 	z^2 \left(T(z,x)^2-1\right) = r(x)\left(T(z,x+\epsilon)+T(z,x)\right)\left(T(z,x-\epsilon)+T(z,x)\right),
\end{equation}	
\begin{equation}
	\label{reccle2}
	\fl	
	2 z R(z,x) = T(z,x+\epsilon)+T(z,x),
\end{equation} 
where for simplicity we do not indicate the dependence of $(r,R,T)$ on $\epsilon$.
These identities  imply
\begin{equation}
	\label{rrrr}
	\fl
	\lambda  (T(z,x)^2-1) = r(x) \left(T(z,x+\epsilon)+T(z,x)\right)\left(T(z,x-\epsilon)+T(z,x)\right).
\end{equation}

We will next consider the exactly solvable linear and double Penner models. By comparing their
continuum limit expansions with direct asymptotic expansions of the exact free energies
we not only check the previous method but are able to obtain asymptotic expansions
for $F_{\rm lin}$. Finally, in section~\ref{sec:cp} we study the cubic Penner model as a nontrivial
example of how to derive the asymptotic expansion for the free energy in a case that is not exactly
solvable.
%%%%%%%%%%%%%%%%%%%%%%%%%%%%%%%%%%%%%%%%%%%%%%%%%%%%%%%%%%%%%%%%%
\subsection{The linear Penner model}
%%%%%%%%%%%%%%%%%%%%%%%%%%%%%%%%%%%%%%%%%%%%%%%%%%%%%%%%%%%%%%%%%
The explicit expressions~(\ref{rp})--(\ref{sp}) of the recurrence coefficients for the linear Penner model shows that its
continuum limit is
\begin{equation}
	r(\epsilon,x)=x+x^2=x(x+1),\quad s(\epsilon,x)=1+2x+\epsilon.
\end{equation}
Substituting the expression for $r(\epsilon,x)$ into equation~(\ref{expfcal}) for the free energy expansion and using~(\ref{axpb}) with
$a=b=1$ we find
\begin{equation}
	\label{flpcl}
	\mathcal{F}(\epsilon,x)
	 =
	\mathcal{F}^G(\epsilon,x)+\mathcal{F}^G(\epsilon,x+1)
	+ \frac{1}{\epsilon^2} \left( x+ \frac{3}{4}\right).
\end{equation}
On the other hand from the exact expression for the partition function~(\ref{pz}) it follows that
\begin{equation}
	\label{flpex}
	\fl
	F_n \sim F
	 \approx
	F^G(\epsilon,x) + F^G(\epsilon,x+1) - \frac{\log\epsilon}{2\epsilon^2} -\frac{1}{2\epsilon}(2x+1)\log(2\pi)-\log G\left(1+\frac{1}{\epsilon}\right).
\end{equation}
Using~(\ref{aba}), (\ref{gau}) and~(\ref{mcfg}) it is easy to check that the asymptotic expansions~(\ref{flpcl})
and~(\ref{flpex}) are in agreement given the following asymptotic expansion for the linear term:
\begin{equation}
	F_{{\rm lin}}
	=
	-\frac{x}{\epsilon^2} + \frac{x}{\epsilon}\log 2\pi
	+\frac{\log\epsilon}{12}+\zeta'(-1)-\sum_{k=1}^{\infty}\frac{\epsilon^{2k}B_{2k+2}}{2k(2k+2)}.
\end{equation}
%%%%%%%%%%%%%%%%%%%%%%%%%%%%%%%%%%%%%%%%%%%%%%%%%%%%%%%%%%%%%%%%%
\subsection{The double Penner model}
Recalling that $\alpha_0=\mu_0 N$ and $\alpha_1=\mu_1 N$, we find that the continuum limit of the recurrence
coefficients~(\ref{eq:dprn})--(\ref{eq:dpsn})  for the double Penner model~(\ref{dps}) are
\begin{equation}
	r(\epsilon,x)
	=
	\frac{x\left(x+\mu _0\right) \left(x+\mu _1\right)\left(x+\mu _0+\mu _1\right)}{
	\left(2x+\mu _0+\mu _1\right){}^2 \left(2x-\epsilon+\mu _0+\mu _1\right)\left(2x+\epsilon+\mu _0+\mu _1\right)},
\end{equation}

\begin{equation}
	s(\epsilon,x)= \frac{2x^2+2x(\mu_0+\mu_1+\epsilon)+(\mu_0+\mu_1)(\mu_0+\epsilon)}{
	                  (2x+\mu_0+\mu_1)(2x+\mu_0+\mu_1+2\epsilon)}.
\end{equation}
The logarithm of the numerator of the expression for $r(\epsilon,x)$ can be directly substituted into~(\ref{expfcal}) and gives four terms
of the type~(\ref{axpb}) to $\mathcal{F}(\epsilon,x)$. However, two of the factors in the denominator
contain the parameter $\epsilon$, and if we apply the same procedure the result would have to be
reexpanded in $\epsilon$. A more efficient approach consists in writing the logarithm of the denominator
as
\begin{eqnarray}
	\log\left(\left(2x+\mu _0+\mu _1\right){}^2 \left(2x-\epsilon+\mu _0+\mu _1\right)\left(2x+\epsilon+\mu _0+\mu _1\right)\right)
	=
	\nonumber\\
	\quad 4\cosh^2\left(\frac{\epsilon}{4}\partial_x\right)\log(2x+\mu_0+\mu_1).
\end{eqnarray}
Substituting this expression into the right hand side of~(\ref{e1}) we find that the contribution $F_\mathrm{d}(\epsilon,x)$
of the denominator satisfies
\begin{equation}
	4 \sinh^2\left(\frac{\epsilon}{2} \partial_x\right) F_\mathrm{d}(\epsilon,x)
	=
	4\cosh^2\left(\frac{\epsilon}{4}\partial_x\right)\log(2x+\mu_0+\mu_1),
\end{equation}
and recalling that $\sinh2x=2\sinh x\cosh x$, we arrive at the following equation:
\begin{equation}
	4 \sinh^2\left(\frac{\epsilon}{4} \partial_x\right)  F_\mathrm{d} = \log(2x+\mu_0+\mu_1).
\end{equation}
I.~e., the whole denominator gives a contribution to $\mathcal{F}(\epsilon,x)$ of the same type~(\ref{axpb}) as
the contributions of each factor in the numerator but with opposite sign, with $\epsilon$
replaced by $\epsilon/2$, $a=2$ and $b=\mu_0+\mu_1$. Summing up, we have
\begin{eqnarray}
 \fl
 \mathcal{F}(\epsilon,x) & = &   \mathcal{F}^G(\epsilon,x)
                                         + \mathcal{F}^G(\epsilon,x+\mu_0)
                                         + \mathcal{F}^G(\epsilon,x+\mu_1)
                                         + \mathcal{F}^G(\epsilon,x+\mu_0+\mu_1)  \nonumber\\
\fl                                      &  & {}- \mathcal{F}^G(\epsilon,2x+\mu_0+\mu_1)\nonumber\\
\fl                                      &  & {} + \frac{1}{4\epsilon^2}
                                             \Big[                                
                                                       \mu _0 \left(3 \mu _0-2  \left(\mu _0+2 x\right)   \log\mu _0 +4 x\right)
                                                       + \mu _1 \left(3 \mu _1-2 \left(\mu _1+2 x\right)\log\mu _1 +4 x\right) \nonumber\\
\fl                                      &  &      \qquad {}+ \left(\mu _0+\mu _1\right) \left(3 \left(\mu _0+\mu _1\right)- 2 \left(\mu _0+\mu _1+2 x\right)\log \left(\mu _0+\mu _1\right)+4 x\right)
                                                   \nonumber\\
\fl                                      &   &   \qquad {}-\left(\mu _0+\mu _1\right) \left(3 \left(\mu _0+\mu _1\right)-2 
                                                    \left(\mu _0+\mu _1+4 x\right)\log \left(\mu _0+\mu _1\right)+8 x\right)\Big],
\end{eqnarray}
where the last bracket is the sum of the last terms in equation~(\ref{axpb}).

On the other hand, from the exact form~(\ref{doup}) of the partition function  we obtain
\begin{eqnarray}
\fl F_n
	&\approx&
	\frac{1}{12}\log\epsilon+\frac{x}{\epsilon}\log 2\pi+\frac{1}{2\epsilon^2}\Big[x^2\log x+(x+\mu_0)^2\log(x+\mu_0)+(x+\mu_1)^2\log(x+\mu_1)
	\nonumber\\
\fl   & &
	{}+(x+\mu_0+\mu_1)^2\log(x+\mu_0+\mu_1)-(2x+\mu_0+\mu_1)^2\log(2x+\mu_0+\mu_1)-\mu_0^2\log\mu_0
	\nonumber\\
\fl   & &
	{}-\mu_1^2\log\mu_1\Big]-\frac{1}{12}\log\frac{x(x+\mu_0)(x+\mu_1)(x+\mu_0+\mu_1)}{(2x+\mu_0+\mu_1)\mu_0\mu_1}+\zeta'(-1)
	\nonumber\\
\fl   & &
	{}+\sum_{k=1}^\infty \frac{\epsilon^{2k}B_{2k+2}}{2k(2k+2)}
        \left[\frac{1}{x^{2k}}+\frac{1}{(x+\mu_0)^{2k}}+\frac{1}{(x+\mu_1)^{2k}}+\frac{1}{(x+\mu_0+\mu_1)^{2k}}\right.
     \nonumber\\
\fl   & &\quad\left.-\frac{1}{(2x+\mu_0+\mu_1)^{2k}}-\frac{1}{\mu_0^{2k}}-\frac{1}{\mu_1^{2k}}\right].
\end{eqnarray}
Again, by comparing the last two equations we find an asymptotic expansion for the linear term,
\begin{eqnarray}
\fl	F_{{\rm lin}}
	&=&
	\frac{1}{12}\log\epsilon+\frac{x}{\epsilon}\log 2\pi
	+\frac{x}{\epsilon^2}\left[\mu_0\log\mu_0+\mu_1\log\mu_1-(\mu_0+\mu_1)\log(\mu_0+\mu_1)\right]
	\nonumber\\
\fl        & &
        {}+\frac{1}{12}\log(\mu_0\mu_1)+\zeta'(-1)
        -\sum_{k=1}^{\infty}\frac{\epsilon^{2k}B_{2k+2}}{2k(2k+2)}\left(\frac{1}{\mu_0^{2k}}+\frac{1}{\mu_1^{2k}}\right).
\end{eqnarray}
%%%%%%%%%%%%%%%%%%%%%%%%%%%%%%%%%%%%%%%%%%%%%%%%%%%%%%%%%%%%%%%%%
\subsection{The  cubic Penner model\label{sec:cp}}
%%%%%%%%%%%%%%%%%%%%%%%%%%%%%%%%%%%%%%%%%%%%%%%%%%%%%%%%%%%%%%%%%
In this section we show how to implement an order-by-order perturbative calculation of the continuum limit
of the free energy for the non exactly solvable cubic Penner model,
\begin{equation}
	W(z)=\frac{z^3}{3}-\log z.
\end{equation}
In this case the $k=0$ string equations~(\ref{str1k})--(\ref{str12k}) reduce to
\begin{equation}
	-1+4{\rho}_0{\sigma}_0-\frac{{\sigma}_0}{\sqrt{{\sigma}_0^2-4{\rho}_0}}=2x,
\end{equation}
\begin{equation}
	2{\rho}_0+{\sigma}_0^2+\frac{1}{\sqrt{{\sigma}_0^2-4{\rho}_0}}=0.
\end{equation}
It follows immediately from this system that $\rho_0$ can be written in terms of $\sigma_0$,
\begin{equation}
	\label{eq:r0cp}
	{\rho}_0=\frac{1+2x-{\sigma}_0^3}{6{\sigma}_0},
\end{equation}
while ${\sigma}_0^3$ satisfies the cubic equation
\begin{equation}
	(1+2x+2{\sigma}_0^3)^2(5{\sigma}_0^3-2(1+2x))
	=
	27{\sigma}_0^3.
\end{equation}
Therefore
\begin{equation}
	\label{eq:s03}
	\sigma_0^3 = \frac{1}{10}\left( -2 (1+2x) + \frac{3(5+(1+2x)^2)}{\Delta^{1/3}} + 3\Delta^{1/3} \right),
\end{equation}
where
\begin{equation}
	\Delta = (2 x+1)^3-5 (2 x+1)+5 \sqrt{-(2 x+1)^4-2(2 x+1)^2-5},
\end{equation}
and where the square and cubic roots are assumed to take their respective principal values. Note that
despite the appearance of intermediate formally complex results, with these determinations the
final result is real. In fact, from~(\ref{eq:s03}) we can find the series expansion of $\sigma_0(x)$
around $x=0$, whose first three terms are
\begin{equation}
	{\sigma}_0(x) = 1+\frac{2}{9}x^2-\frac{4}{81}x^3+\cdots.
\end{equation}
Using~(\ref{eq:r0cp}) we get
\begin{equation}
	{\rho}_0(x) = \frac{x}{3}-\frac{x^2}{9}-\frac{4}{81}x^3+\cdots,
\end{equation}
and from~(\ref{efes0}) we find that
\begin{equation}
	\mathcal{F}_0(x) = \frac{x^2}{2}\log x-\frac{3+\log9}{4}x^2-\frac{x^3}{18}+\cdots.
\end{equation}

Proceeding to the next order, from~(\ref{cons}) in appendix~B we have that
\begin{equation}
	{\sigma}_1(x)={\sigma}'_0(x)/2.
\end{equation}
The system to calculate the next terms, although rather unwieldy, has a simple structure:
it is a nonhomogeneous linear system of equations for $\rho_1$ and $\sigma_2$ whose coefficients and
independent terms can be expressed as functions of $\sigma_0$, $\rho_0$ and their
derivatives (or, using~(\ref{eq:r0cp}), as functions of $\sigma_0$ and $x$). We write it explicitly to
show the pattern. Setting $k=1$ in the string equations (\ref{str1k})--(\ref{str12k}) we find,
\begin{equation}
	M \left[\begin{array}{c} \rho_1 \\ \sigma_2 \end{array}\right]
	=
	 \left[\begin{array}{c} \delta_1 \\ \delta_2 \end{array}\right],
\end{equation}
where the symmetric $2\times 2$ matrix $M$ is given by
\begin{equation}
	M_{11} = M_{22} = 4{\sigma}_0\left(2({\sigma}_0^2-4{\rho}_0)^{7/2}-({\sigma}_0^2-4{\rho}_0)^2\right),
\end{equation}
\begin{equation}
	M_{12} = M_{21} = 8{\rho}_0\left(({\sigma}_0^2-4{\rho}_0)^{7/2}+({\sigma}_0^2-4{\rho}_0)^2\right),
\end{equation}
and where the independent terms are,
\begin{eqnarray}
	\delta_1
	=
	{\rho}_0\Big[4{\sigma}_0\left(5({\rho}_0')^2+4{\rho}_0(({\sigma}_0')^2-{\rho}_0'')\right)
	+{\sigma}_0^3(({\sigma}_0')^2+4{\rho}_0'')
	\nonumber\\
	\quad{}-8{\sigma}_0^2(2{\rho}_0'{\sigma}_0'+{\rho}_0{\sigma}_0'')
	           +16{\rho}_0(2{\rho}_0{\sigma}_0''-{\rho}_0'{\sigma}_0')\Big],
\end{eqnarray}
and
\begin{eqnarray}
	\fl
	\delta_2 = -\Big[({\sigma}_0^2-4{\rho}_0)\left(12({\rho}_0')^2-10{\sigma}_0{\rho}_0'{\sigma}_0'+
                         ({\sigma}_0^2-4{\rho}_0)\left(1+({\sigma}_0^2-4{\rho}_0)^{3/2}\right)(({\sigma}_0')^2+2{\rho}_0'')\right)\nonumber\\
        \fl
	\quad{}+40{\rho}_0^2({\sigma}_0')^2
	+2{\rho}_0\left(20({\rho}_0')^2-20{\sigma}_0{\rho}_0'{\sigma}_0'-({\sigma}_0^2-4{\rho}_0)(-7({\sigma}_0')^2-4{\rho}_0''+
			2{\sigma}_0{\sigma}_0'')\right)\Big].
\end{eqnarray}
The system can be easily solved, and we find the expansions
\begin{equation}
	{\rho}_1(x) = -\frac{x}{27}+\frac{23}{243}x^2+\frac{128}{729}x^3+\cdots,
\end{equation}
\begin{equation}
	{\sigma}_2(x) = \frac{1}{9}-\frac{2}{81}x-\frac{52}{81}x^2+\frac{968}{2187}x^3+\cdots.
\end{equation}
Finally, using~(\ref{efes1}) and term-by-term integration we can calculate as many terms as desired of the series
for the next order (in $\epsilon$), which up to $\mathcal{O}(x^4)$ turns out to be
\begin{equation}
	\mathcal{F}_1(x)=-\frac{1}{12}\log x+\frac{\log 3}{12}+\frac{x}{36}-\frac{25}{648}x^2+\frac{7}{324}x^3+\cdots.
\end{equation}
Higher order terms quickly get more complicated but, once the structure is known, the whole procedure can be
easily programmed in a symbolic computation environment.
%%%%%%%%%%%%%%%%%%%%%%%%%%%%%%%%%%%%%%%%%%%%%%%%%%%%%%%%%%%%%%%%%
%% LARGE N TWO CUT %%%%%%%%%%%%%%%%%%%%%%%%%%%%%%%%%%%%%%%%%%%%%%%%%%%%
%%%%%%%%%%%%%%%%%%%%%%%%%%%%%%%%%%%%%%%%%%%%%%%%%%%%%%%%%%%%%%%%%
\section{The continuum limit for $Z_2$-symmetric models in the two-cut case\label{sec:lnlz2}}
%%%%%%%%%%%%%%%%%%%%%%%%%%%%%%%%%%%%%%%%%%%%%%%%%%%%%%%%%%%%%%%%%
In this section we apply the method of orthogonal polynomials to calculate the continuum limit of $Z_2$-symmetric
Penner models in the two-cut case.

The recurrence coefficient $s_n$ for $Z_2$-symmetric models is identically zero, and from the experience with matrix
models associated to polynomial potentials~\cite{TA92} it is natural to assume the appearance of two different expansions
for $r_n$ (as well as for $F_n$) according to the parity of $n$. More concretely, the analog of~(\ref{main}) is the pair of equations
\begin{equation}
	\label{mains}
	\frac{Z_{2n+3}  Z_{2n-1}}{Z_{2n+1}^2}=r_{2n+2}r_{2n+1}^2 r_{2n},
	\quad
	\frac{Z_{2n+2}  Z_{2n-2}}{Z_{2n}^2}=r_{2n+1}r_{2n}^2 r_{2n-1},
\end{equation}
where three successive even and odd terms of the partition function are written separately
as products of recurrence coefficients $r_n$. We assume that the continuum limit of the recurrence coefficient $r_n$
in the two-cut case is represented by a two-branch asymptotic expansion
\begin{equation}
	\label{2b}
	r_n \sim
	\left\{
		\begin{array}{cc}
			\displaystyle
			a(\epsilon,X)\approx\sum_{k=0}^\infty \epsilon^{2k}   \alpha_k(X),& \mbox{ for $n$ odd,} \\
			\displaystyle
		 	b(\epsilon,X)\approx\sum_{k=0}^\infty \epsilon^{2k}   \beta_k(X), & \mbox{for $n$ even,}
		\end{array}
	\right.
\end{equation}
where
\begin{equation}
	X = \frac{n}{N}, \quad \epsilon = \frac{1}{N},
\end{equation}
and the coefficients $\alpha_k$ and $\beta_k$ are analytic functions of $X$ in a certain real interval.
Thus we have
\begin{equation}
	\label{r2}
	r_{2n+1} \sim a(\epsilon,x+\epsilon),\quad 
	r_{2n} \sim b(\epsilon,x),\quad
\end{equation}
where
\begin{equation}
	x = 2 X = \frac{2n}{N}
\end{equation}
and
\begin{equation}
	\label{r3}
	a(\epsilon,x)
	\approx
	\sum_{k=0}^\infty\alpha_k(x) \epsilon^{2k},\quad
	b(\epsilon,x)
	\approx
	\sum_{k=0}^\infty\beta_k(x) \epsilon^{2k}.
\end{equation}
Likewise, we assume that in the continuum limit there exists a two-branch representation
for the free energy
\begin{equation}
	\label{2bf0}
	F_n \sim \left\{
	                       \begin{array}{cc}
	                       		A(\epsilon,X),& \mbox{ for $n$ odd,} \\
	                       		B(\epsilon,X), & \mbox{for $n$ even,}
				\end{array}
			\right.
\end{equation}
or, equivalently, that we have
\begin{equation}
	\label{rf2}
	F_{2n+1} \sim A(\epsilon,x+\epsilon),\quad 
	F_{2n} \sim B(\epsilon,x).
\end{equation}
In this way, the identities~(\ref{mains}) lead to
\begin{equation}
	\label{mains2}
	\left\{\begin{array}{c}
		A(x+2\epsilon)+A(x-2\epsilon)-2A(x)=\log\left(b(x+\epsilon)b(x-\epsilon)a(x)^2\right),\\
		B(x+2\epsilon)+B(x-2\epsilon)-2B(x)=\log\left(a(x+\epsilon)a(x-\epsilon)b(x)^2\right),
	\end{array}\right.
\end{equation}
where, for simplicity, we do not indicate the dependence on $\epsilon$. Using~(\ref{idb}) we find
\begin{equation}
	\everymath{\displaystyle}
		\label{mains3}
		\left\{\begin{array}{c}
			4 \epsilon^2 \partial_{xx} A(x)
			=
			\mathrm{L}(2 \epsilon \partial_x) \log\left(b(x+\epsilon)b(x-\epsilon)a(x)^2\right),\\
			4 \epsilon^2 \partial_{xx} B(x)
			=
			\mathrm{L}(2 \epsilon \partial_x) \log\left(a(x+\epsilon)a(x-\epsilon)b(x)^2\right).
		\end{array}\right.
\end{equation}
where $\mathrm{L}(\epsilon\partial_x)$ is the infinite-order differential operator defined in  (\ref{ope}).
Therefore, there are  decompositions of $A(\epsilon,x)$ and $B(\epsilon,x)$  of the form 
\begin{eqnarray}
	\label{AB1}
	\nonumber
	A(\epsilon,x) & = & A_{{\rm lin}}+ \frac{1}{4\epsilon^2}\int_0^x (x-t)\log\left(b(t+\epsilon)b(t-\epsilon)a(t)^2\right)  \rmd  t \\
	\nonumber \\
        &  & {}-\sum_{k=1}^{\infty} \frac{B_{2k}(2\epsilon\partial_x)^{2k-2}}{(2k) (2k-2)!} 
                   \log\left(b(x+\epsilon)b(x-\epsilon)a(x)^2\right),
\end{eqnarray}
\begin{eqnarray}
	\label{AB2}
	B(\epsilon,x) & = & B_{{\rm lin}}+ \frac{1}{4\epsilon^2}\int_0^x (x-t)\log\left(a(t+\epsilon)a(t-\epsilon)b(t)^2\right)  \rmd  t 
	\nonumber \\
        &  & {}-\sum_{k=1}^{\infty} \frac{B_{2k}(2\epsilon\partial_x)^{2k-2}}{(2k) (2k-2)!} 
                   \log\left(a(x+\epsilon)a(x-\epsilon)b(x)^2\right),
\end{eqnarray}
where $A_{{\rm lin}}$ and $B_{{\rm lin}}$ are $\epsilon$-dependent linear terms in $x$.
Then, using the expansions~(\ref{r3}) we deduce the existence of asymptotic expansions
in powers of $\epsilon^2$ of the form
\begin{equation}
	\label{expf2ab}
	A(\epsilon,x) - A_{{\rm lin}}
	\approx  \sum_{k=0}^\infty \epsilon^{2k-2} \mathcal{A}_k(x),
	\quad
	B(\epsilon,x) - B_{{\rm lin}}
	\approx  \sum_{k=0}^\infty \epsilon^{2k-2} \mathcal{B}_k(x),
\end{equation}
For example, the first few terms are 
\begin{equation}
	\label{0} 
\fl
	\mathcal{A}_0
	=
	\mathcal{B}_0
	=\frac{1}{2}\int_0^x (x-t)\log \left(\alpha_0(t) \beta_0(t)\right)  \rmd  t,
\end{equation}
\begin{equation}
	\label{1}
\fl
	\mathcal{A}_1
	=
	\frac{1}{2}\int_0^x (x-t)
	\left(\frac{\beta_1(t)}{\beta_0(t)}+\frac{\alpha_1(t)}{\alpha_0(t)}
		+\frac{\beta_0''(t)}{2\beta_0(t)}-\frac{1}{2}\left(\frac{\beta_0'(t)}{\beta_0(t)}\right)^2\right)  \rmd  t
	-\frac{1}{6} \log\left(\alpha_0\beta_0\right),
\end{equation}
\begin{equation}
	\label{2}
\fl
 	\mathcal{B}_1
	=
	\frac{1}{2}\int_0^x (x-t)
	\left(\frac{\beta_1(t)}{\beta_0(t)}+\frac{\alpha_1(t)}{\alpha_0(t)}
		+\frac{\alpha_0''(t)}{2\alpha_0(t)}-\frac{1}{2}\left(\frac{\alpha_0'(t)}{\alpha_0(t)}\right)^2\right)  \rmd  t
	-\frac{1}{6} \log\left(\alpha_0 \beta_0\right).
\end{equation}
As in the one-cut case, the explicit form of  $A_{{\rm lin}}$ and $B_{{\rm lin}}$ for non exactly solvable
models  can be obtained  applying the Euler-MacLaurin summation formula  to $F_{2n+1}$ and to $F_{2n}$
respectively. However, since $b(\epsilon,x)$ vanishes near $x=0$ it is required to regularize $F_{2n}$. 

We next describe the two-cut version of  our method to determine the coefficients $\alpha_k$ and $\beta_k$
of the expansions~(\ref{r3}) from the continuum limit of the string equations and the resolvent  identities.
%%%%%%%%%%%%%%%%%%%%%%%%%%%%%%%%%%%%%%%%%%%%%%%%%%%%%%%%%%%%%%%%%
\subsection{Solutions of the string equations in the continuum limit}
%%%%%%%%%%%%%%%%%%%%%%%%%%%%%%%%%%%%%%%%%%%%%%%%%%%%%%%%%%%%%%%%%
Following the pattern of a previous analysis valid for matrix models with polynomial potentials~\cite{ALA11},
in order to perform the continuum limit of  the string and recurrence equations~(\ref{stre1}) and~(\ref{recce1})
we introduce two asymptotic expansions for the continuum limits $U$ and $V$ of the odd and even $T_n$,
\begin{equation}
     T_{2 n+1}(\lambda)\sim U(\epsilon,\lambda,x) \approx \sum_{k=0}^\infty U_k(\lambda,x) \epsilon^{2k},
\end{equation}
\begin{equation}
     T_{2 n}(\lambda)\sim V(\epsilon,\lambda,x) \approx \sum_{k=0}^\infty V_k(\lambda,x) \epsilon^{2k}.
\end{equation}
Then the continuum limit of the relation~(\ref{recce1}) between $T_n(z)$ and $r_n$ is equivalent
to the following system:
\begin{equation}
 	\label{resc2a}
\fl
  	a(x)   \left(U(\lambda,x)+ V(\lambda,x-\epsilon)\right) \left(U(\lambda,x)+ V(\lambda,x+\epsilon)\right)
	=
	\lambda  \left(U(\lambda,x)^2-1\right),
\end{equation}
\begin{equation}
	\label{resc2b}
\fl
    	b(x)  \left(V(\lambda,x)+U(\lambda,x-\epsilon)\right) \left( V(\lambda,x)+U(\lambda,x+\epsilon)\right)
	=
	\lambda \left(V(\lambda,x)^2-1\right),
\end{equation}
where for simplicity the dependence on $\epsilon$ is not indicated. Identification of the coefficients of $\epsilon^{2k}$
in~(\ref{resc2a})--(\ref{resc2b}) leads
to a system of recurrence relations for the  coefficients $U_k$ and $V_k$ (see appendix~B). 
For example, for $k=0$ we find that
\begin{equation}\label{UV0}
 U_0 = (\alpha_0-\beta_0+\lambda)u(\lambda,x),\quad
 V_0 = (\beta_0-\alpha_0+\lambda)u(\lambda,x),
\end{equation}
where
\begin{equation}
 u(\lambda,x)=\frac{1}{\sqrt{ \left(\lambda-(\alpha_0+\beta_0)\right)^2-4\alpha_0\beta_0}}.
\end{equation}

In this way the continuum  limit of the string equation~(\ref{stre1})  splits into the
system of equations
\begin{equation}
	\everymath{\displaystyle}
	\label{stre1v}
	\left\{\begin{array}{l}
		\oint_{\Gamma_{\infty}}\frac{\rm{d} \lambda}{2\pi i}W_0'(\lambda) U(\epsilon,\lambda,x)
		+
		\sum_{i=1}^k \mu_i \left(U(\epsilon,q_i^2,x)-1\right)=x,\\
 		\oint_{\Gamma_{\infty}}\frac{\rm{d} \lambda}{2\pi i}W_0'(\lambda) V(\epsilon,\lambda,x)
		+
		\sum_{i=1}^k \mu_i \left(V(\epsilon,q_i^2,x)-1\right)=x.
	 \end{array}\right.
\end{equation}
Again, by identifying the coefficients of $\epsilon^{2k}$ in this system we can determine recursively the
coefficients $\alpha_k(x)$ and $\beta_k(x)$. For example, using~(\ref{UV0}) and setting $k=0$ in
the system~(\ref{stre1v}), we obtain the following  equations  for the leading coefficients
$\alpha_0(x)$ and $\beta_0(x)$:
\begin{equation}
	\label{sysrse0}
\fl		\frac{1}{2\pi\rmi}\oint_{\Gamma_{\infty}}
		(\lambda+\alpha_0-\beta_0)u(\lambda,x)W_0'(\lambda)\rmd \lambda
		+
		\sum_{i=1}^m \mu_i \left((q_i^2+\alpha_0-\beta_0)u(q_i^2,x)-1\right)
		=
		x,
\end{equation}
\begin{equation}
	\label{sysrse2}	
\fl		\frac{1}{2\pi\rmi}\oint_{\Gamma_{\infty}}
		(\lambda+\beta_0-\alpha_0)u(\lambda,x)W_0'(\lambda)\rmd \lambda
		+
		\sum_{i=1}^m \mu_i \left((q_i^2+\beta_0-\alpha_0)u(q_i^2,x)-1\right)
		=
		x.		
\end{equation}

As a noteworthy application, these equations can be used to calculate the endpoints of the 2-cut
eigenvalue support $[-b,-a] \cup [a,b]$, since $a$ and $b$ are related to $\alpha_0$ and $\beta_0$
according to by
\begin{equation}
	\label{1c}
	a = \sqrt{\alpha_0}-\sqrt{\beta_0},
	\quad
	b = \sqrt{\alpha_0}+\sqrt{\beta_0}.
\end{equation}
(See~\cite{TA92} for Penner models and~\cite{{ALA11}} for polynomial potentials.)
%%%%%%%%%%%%%%%%%%%%%%%%%%%%%%%%%%%%%%%%%%%%%%%%%%%%%%%%%%%%%%%%%
\subsection{The gaussian Penner model}
%%%%%%%%%%%%%%%%%%%%%%%%%%%%%%%%%%%%%%%%%%%%%%%%%%%%%%%%%%%%%%%%%
As an illustration of the previous formalism we will now use the exact expression for the partition function~(\ref{dpz})
of the  gaussian Penner model to confirm the existence of the two-branch structure for the free energy,
\begin{equation}
	\everymath{\displaystyle}
	\label{2bf}
	F_n \sim
	\left\{\begin{array}{cc} 
		A(\epsilon,X)\approx A_{{\rm lin}}+\sum_{k=0}^\infty \epsilon^{2k-2}  \mathcal{A}_k(X),& \mbox{ for $n$ odd,} \\
		B(\epsilon,X)\approx B_{{\rm lin}}+\sum_{k=0}^\infty \epsilon^{2k-2}  \mathcal{B}_k(X), & \mbox{for $n$ even,}
	\end{array}\right.
\end{equation}
and to check our formulas~(\ref{0})--(\ref{2}).  

If to the effect of asymptotic expansion we consider as constants the bounded terms $(-1)^n$, it is straightforward
to use the asymptotic expansion~(\ref{aba}) of the Barnes function in~(\ref{dpz}) and compute the first terms of
the continuum limit of the free energy for the gaussian Penner model:
\begin{eqnarray}
	\label{exp30}
	\nonumber
	\fl
	F_n
	\approx
	\frac{1}{4 \epsilon^2} \left(X^2 \log X+(1+X) (-3 X+(1+X) \log(1+X) \right)
	+\frac{\log (2\pi)}{\epsilon} X+ \frac{\log \epsilon}{6} 
	\nonumberÊ\\
	\fl
	\quad{}+\frac{1}{6}\Big(\log 2+12 \zeta'(-1)-\frac{1}{4}\Big((1+3(-1)^n)\log X+(1-3(-1)^n)\log (1+X) \Big)\Big)
	\nonumber \\
	\fl
	\quad{}-\frac{\epsilon ^2}{480} \left(\frac{15 (-1)^n+1}{X^2}+\frac{1-15 (-1)^n}{(X+1)^2}+14\right)+\mathcal{O}(\epsilon^4).
\end{eqnarray}
By separating the even and the odd terms in~(\ref{exp30}) we can show explicitly that the two branches
of the continuum limit of the free energy~(\ref{2bf}) are given by
\begin{eqnarray}
	\label{expB0}
	\fl 
	A(\epsilon,X) \approx \frac{1}{4 \epsilon^2} \left(X^2 \log X+(1+X) (-3 X+(1+X) \log(1+X) \right) \nonumber \\
 	\fl
	\quad{}+\frac{\log (2\pi)}{\epsilon} X+ \frac{\log \epsilon}{6}
	           +\frac{1}{6}\left(\log 2+12 \zeta'(-1)\right)+\frac{1}{12}\log\left(\frac{X}{(1+X)^2}\right) 
	           +\mathcal{O}(\epsilon^2),
\end{eqnarray}
and
\begin{eqnarray}
	\label{expA0}
	\fl
	B(\epsilon,X)\approx \frac{1}{4 \epsilon^2} \left(X^2 \log X+(1+X) (-3 X+(1+X) \log(1+X) \right) \nonumber\\
	\fl
	\quad{}+\frac{\log (2\pi)}{\epsilon} X+ \frac{\log \epsilon}{6}
	+\frac{1}{6}\left(\log 2+12 \zeta'(-1)\right)
	+\frac{1}{12}\log\left(\frac{1+X}{X^2}\right)+\mathcal{O}(\epsilon^2).
\end{eqnarray}
Moreover, these expansions are in complete agreement with~(\ref{0})--(\ref{2}), since
according to (\ref{rgs}) 
\begin{equation}
	\label{gp1}
	a(\epsilon,X)=\alpha_0(X)=X+1,
	\quad
	b(\epsilon,X)=\beta_0(X)=X.
\end{equation}
%%%%%%%%%%%%%%%%%%%%%%%%%%%%%%%%%%%%%%%%%%%%%%%%%%%%%%%%%%%%%%%%%%%%%%
\begin{figure}
    \begin{center}
        \includegraphics[width=12cm]{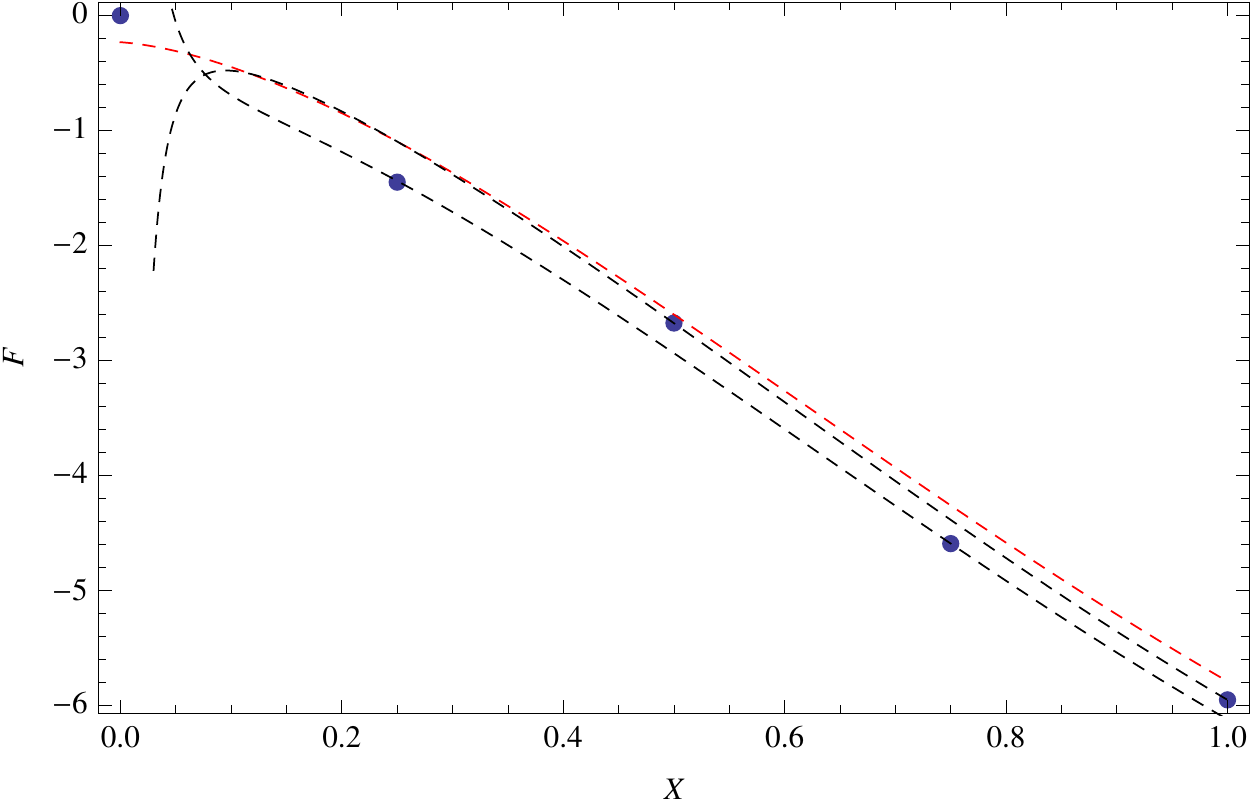}
    \end{center}
    \caption{Exact values (dots) and asymptotic formulas for the continuum limit of the free energy
                 of the Gaussian-Penner matrix model with $N=4$: the red dashed line represents the terms
                 of~(\ref{exp30}) up to $\mathcal{O}(1)$, while the black dashed lines
                 represent the branches $A(1/4,X)$ and $B(1/4,X)$ up to $\mathcal{O}(\epsilon^4)$.\label{fig:gp}}
\end{figure}
%%%%%%%%%%%%%%%%%%%%%%%%%%%%%%%%%%%%%%%%%%%%%%%%%%%%%%%%%%%%%%%%%%%%%%
It is important to notice the different behavior of the two branches of the free energy in the continuum limit
\begin{equation}\label{disc}
A(\epsilon,X)-B(\epsilon,X)=\frac{1}{4} \log\left(\frac{X}{X+1} \right)+\mathcal{O}\left(\epsilon^2\right).
\end{equation}

In figure~\ref{fig:gp} we illustrate these results with a numerical calculation. The dots are the exact values
of $F_n=\log Z_n$ obtained from the exact partition function~(\ref{dpz}) with $N=4$ and $n=0,\ldots,4$.
The red dashed line is the common part up to $\mathcal{O}(1)$ of~(\ref{exp30}) (i.e., the three terms
in the first row), and the two black dashed lines represent the odd branch $A(1/4,X)$ and even branch
$B(1/4,X)$ up to $\mathcal{O}(\epsilon^4)$. Note how, to the given precision, the exact values fall
alternately on the corresponding branch of the expansion.

Finally, we mention a very general result first derived heuristically in~\cite{BO00,EY09} and further developed
in~\cite{BO12}, whereby the free energy of matrix models with a disconnected eigenvalue support
involves the logarithm of a theta function. According to~\cite{BO00}, in the symmetric two-cut case
(see~(2.45) and~(3.24) in~\cite{BO00}) the corresponding term turns out to be
$\log(\theta_3(n/2))$, where $n$ is the matrix dimension. As a consequence of
the periodicity relation $\theta_3(z+1)=\theta_3(z)$ the theta function takes only the two values
$\theta_3(1/2)$ for odd $n$ and $\theta_3(0)$ for even $n$, which leads to the two-branched structure of
the asympotic expansion for the free energy.
%%%%%%%%%%%%%%%%%%%%%%%%%%%%%%%%%%%%%%%%%%%%%%%%%%%%%%%%%%%%%%%%%
%% CONCLUDING %%%%%%%%%%%%%%%%%%%%%%%%%%%%%%%%%%%%%%%%%%%%%%%%%%%%%%%
%%%%%%%%%%%%%%%%%%%%%%%%%%%%%%%%%%%%%%%%%%%%%%%%%%%%%%%%%%%%%%%%%
\section{Concluding remarks\label{sec:cr}}
In this paper we have  shown how the method of  orthogonal polynomials for Penner models can be applied
to study the partition function of Penner matrix models and to compute the continuum expansion of the free energy.
The  key ingredient in our study is a method to solve the string equations which uses certain identities for the resolvent
of the Jacobi matrix defining the three-term recursion relation for the orthogonal polynomials.

We have applied this method to compute the partition functions of several of the exactly solvable Penner models,
thus providing  an alternative derivation of the exact results obtained via the Selberg's integral.
In addition, we have also shown that in the continuum limit the free energy of certain exactly solvable
models like the linear and double Penner models can be written as a sum of gaussian contributions
plus linear terms.

For non-exactly solvable Penner models we have provided a perturbative method for solving the  system of
string equations and resolvent identities, thus determining the large $N$ expansion  of the free energy. 
Although in this paper we have dealt only with solutions that are asymptotic power series in the
small parameter $\epsilon=1/N$, it has been recently shown~\cite{SC14,PA10,MA08b},  one might
also consider trans-series solutions (i.e., formal series with exponentially small corrections)
that go beyond the usual  large $N$ expansion and describe nonperturbative effects.
Finally, we have discussed and illustrated numerically the double-branch structure of
the free energy for the gaussian Penner model.

Although in the examples shown in this paper we have focused on Penner models defined over paths contained
in the real line (hermitian models), our results hold for Penner models defined on more general paths
(holomorphic nonhermitian models), which have been studied in~\cite{AM94,MA06}.
In these cases, however, a more detailed analysis based on the concept of $S$-curve is required to identify allowable
paths leading to well-defined matrix models~\cite{AL13}.
%%%%%%%%%%%%%%%%%%%%%%%%%%%%%%%%%%%%%%%%%%%%%%%%%%%%%%%%%%%%%%%%%
%% ACK %%%%%%%%%%%%%%%%%%%%%%%%%%%%%%%%%%%%%%%%%%%%%%%%%%%%%%%
%%%%%%%%%%%%%%%%%%%%%%%%%%%%%%%%%%%%%%%%%%%%%%%%%%%%%%%%%%%%%%%%%
\section*{Acknowledgements}
The financial support of the Ministerio de Ciencia e Innovaci\'on under project FIS2011-22566
is gratefully acknowledged.
%%%%%%%%%%%%%%%%%%%%%%%%%%%%%%%%%%%%%%%%%%%%%%%%%%%%%%%%%%%%%%%%%
%% APPENDIX A %%%%%%%%%%%%%%%%%%%%%%%%%%%%%%%%%%%%%%%%%%%%%%%%%%%%%%%%
%%%%%%%%%%%%%%%%%%%%%%%%%%%%%%%%%%%%%%%%%%%%%%%%%%%%%%%%%%%%%%%%%
\section*{Appendix A: Resolvent identities}
In this appendix we prove the resolvent identities (\ref{eq:rid1})--(\ref{eq:rid2}) that allow us to compute $R_n(z)$ and
$T_n(z)$ in terms of the recurrence coefficients.

To prove~(\ref{eq:rid2}) we substitute
\begin{equation}
	(L^k)_{n n} = \frac{1}{h_n}\int_{\gamma}\rme^{-NW(\zeta)}\zeta^k P_n(\zeta)^2  \rmd \zeta
\end{equation}
and
\begin{equation}
	(L^k)_{n n-1} = \frac{1}{h_{n-1}}\int_{\gamma}\rme^{-NW(\zeta)}\zeta^k P_n(\zeta)P_{n-1}(\zeta)  \rmd \zeta,
\end{equation}
into the $z\to\infty$ expansions~(\ref{expr}) and~(\ref{expt}) for $R_n(z)$ and $T_n(z)$, and obtain
\begin{equation}
	\label{Rint}
	R_n(z) = \frac{1}{z}\left(1 + \frac{1}{h_n}\int_{\gamma}\rme^{-NW(\zeta)}\frac{\zeta P_n(\zeta)^2}{z-\zeta}  \rmd \zeta\right),
\end{equation}
and
\begin{equation}
	\label{Tint}
	T_n(z) = 1 + \frac{2}{h_{n-1} z}\int_{\gamma}\rme^{-NW(\zeta)}\frac{\zeta P_n(\zeta)P_{n-1}(\zeta)}{z-\zeta}  \rmd \zeta.
\end{equation}
Now from (\ref{Rint}) and using  (\ref{Tint}) we  deduce  that
\begin{eqnarray}
	\everymath{\displaystyle}
	\fl
	z^2R_n(z) & = &z + \frac{1}{h_n}\left(\int_{\gamma}\rme^{-NW(\zeta)} \zeta P_n(\zeta)^2  \rmd \zeta
	                             + \int_{\gamma}\rme^{-NW(\zeta)}\frac{\zeta^2 P_n(\zeta)^2}{z-\zeta}  \rmd \zeta\right)
	\nonumber\\
	\fl
                          & = &z+s_n+ \frac{1}{h_n}\int_{\gamma}\zeta\,\rme^{-NW(\zeta)}\frac{ P_n(\zeta) P_{n+1}(\zeta)+s_n P_n(\zeta)^2+r_n P_n(\zeta) P_{n-1}(\zeta)}{z-\zeta}  \rmd \zeta
         \nonumber\\ 
         \fl
  			& = &z\,s_n  R_n(z) + \frac{z}{2}\left(T_{n+1}(z)+T_n(z)\right),
\end{eqnarray}
which is the resolvent identity (\ref{eq:rid2}).

To prove~(\ref{eq:rid1}) we use~(\ref{Tint}) and proceed in the form
\begin{eqnarray}
	\everymath{\displaystyle}
	\fl
	z^2\left(T_{n+1}(z)-T_n(z)\right)
	\nonumber\\
	\fl
	= \frac{2}{h_n}
	\left(\int_{\gamma}\rme^{-NW(\zeta)} \zeta P_{n+1}(\zeta) P_n(\zeta) \rmd \zeta + \int \rme^{-NW(\zeta)}\frac{\zeta^2 P_{n+1}(\zeta) P_n(\zeta)}{z-\zeta} \rmd \zeta\right)
	\nonumber\\ 
	\fl
	\quad{}-\frac{2}{h_{n-1}}\left(\int_{\gamma}\rme^{-NW(\zeta)} \zeta P_{n}(\zeta) P_{n-1}(\zeta) \rmd \zeta + \int \rme^{-NW(\zeta)}\frac{\zeta^2 P_{n}(\zeta) P_{n-1}(\zeta)}{z-\zeta} \rmd \zeta\right)
	\nonumber\\
	\fl
        = \frac{2}{h_n}
        \left(h_{n+1} + \int_{\gamma}\zeta\,\rme^{-NW(\zeta)}\frac{P_{n+1}(\zeta)^2+s_n P_{n+1}(\zeta) P_n(\zeta)+r_n  P_{n+1}(\zeta) P_{n-1}(\zeta)}{z-\zeta}  \rmd \zeta\right)
        \nonumber\\
        \fl
	\quad{}-\frac{2}{h_{n-1}}
	\left(h_n + \int_{\gamma}\zeta\,\rme^{-NW(\zeta)}\frac{ P_{n+1}(\zeta) P_{n-1}(\zeta)+s_n P_n(\zeta) P_{n-1}(\zeta)+r_n P_{n-1}(\zeta)^2}{z-\zeta}  \rmd \zeta\right)
	\nonumber\\
	\fl
        =2 z \left(r_{n+1}R_{n+1}(z)-r_nR_{n-1}(z)\right)+z\,s_n  \left(T_{n+1}(z)-T_n(z)\right).
\end{eqnarray}
Thus, we have that
\begin{equation}
	\label{auxrecc}
	(z-s_n)\left(T_{n+1}(z)-T_n(z)\right) = 2\left(r_{n+1}R_{n+1}(z)-r_nR_{n-1}(z)\right).
\end{equation}
Multiplying~(\ref{eq:rid2}) by $T_{n+1}(z)-T_{n}(z)$ and using~(\ref{auxrecc}) we obtain
\begin{equation}
	T_n(z)^2-4r_nR_n(z)R_{n-1}(z)=T_{n+1}(z)^2-4r_{n+1}R_{n+1}(z)R_n(z),
\end{equation}
which means that
\begin{equation}
	T_n(z)^2-4r_nR_n(z)R_{n-1}(z)
\end{equation}
is independent of $n$, and the resolvent identity~(\ref{eq:rid1}) follows from the initial conditions $r_0=0$, $T_0(z)=1$.
%%%%%%%%%%%%%%%%%%%%%%%%%%%%%%%%%%%%%%%%%%%%%%%%%%%%%%%%%%%%%%%%%
%% APPENDIX B %%%%%%%%%%%%%%%%%%%%%%%%%%%%%%%%%%%%%%%%%%%%%%%%%%%%%%%%
%%%%%%%%%%%%%%%%%%%%%%%%%%%%%%%%%%%%%%%%%%%%%%%%%%%%%%%%%%%%%%%%%
\section*{Appendix B: Solutions of the string equations in the continuum limit}
In this appendix we analyze the existence of the expansions~(\ref{exr}) and~(\ref{2b}) of the recurrence
coefficients in the continuum limit.

Let us start with the one-cut case. If we assume expansions of  the form
 \begin{equation}\label{exgo}
	r(\epsilon,x) \approx \sum_{k=0}^{\infty} \epsilon^{ k} {\tilde{\rho}}_k (x),
	\quad
	s(\epsilon,x) 
	                    \approx \sum_{k=0}^{\infty} \epsilon^{k} {\sigma}_k (x),
\end{equation}
 \begin{equation}
 	\label{exgn}
	R(\epsilon,z,x) \approx \sum_{k=0}^{\infty} \epsilon^k R_k (z,x),
	\quad
	T(\epsilon,z,x) \approx \sum_{k=0}^{\infty} \epsilon^{k} \widetilde{T}_k (z,x),
\end{equation}
and identify the coefficients of $\epsilon^k$ in~(\ref{reccl1})--(\ref{reccl2}) we obtain two recursion relations
for the coefficients $R_k$ and $\widetilde{T}_k$:
 \begin{equation}
 	\label{recc11}
	\fl
	\everymath{\displaystyle}
	\begin{array}{l}
	\sum_{i+j=k}\widetilde{T}_i(z,x)\widetilde{T}_j(z,x)
	-
	4\sum_{i+j+l+m=k}\frac{(-1)^i}{i!}\tilde{{\rho}}_m(x)R_l(z,x)\partial_x^iR_j(z,x)=\delta_{k0},\\
	z R_k(z,x)-\sum_{i+j=k}{\sigma}_i(x)R_j(z,x)
	=
	\widetilde{T}_k(z,x)+\frac{1}{2}\sum_{i+j=k, j<k}\frac{1}{i!}\partial_x^i\widetilde{T}_j(z,x).
	\end{array}
\end{equation}
For $k=0$ we have
\begin{equation}
	\label{RT0}
	R_0(z,x) = w(z,x),
	\quad
	\widetilde{T}_0(z,x)=(z-{\sigma}_0(x))w(z,x).
\end{equation}
Using induction it is straightforward to  prove that the coefficients  $R_k$ and $\widetilde{T}_k$ $(k\geq1)$
can be expressed in the form 
\begin{equation}
	\label{RTk}
	\everymath{\displaystyle}
	\begin{array}{l}
		R_k(z,x) = \sum_{j=1}^{2k-1}\left(\alpha_{k,j}(x)+\beta_{k,j}(x)(z-{\sigma}_0(x))\right)w(z,x)^{2j+1},\\
		\widetilde{T}_k(z,x)=\sum_{j=1}^{2k-1}\left(\gamma_{k,j}(x)+\xi_{k,j}(x)(z-{\sigma}_0(x))\right)w(z,x)^{2j+1},
	\end{array}
\end{equation}
where
\begin{equation}
	w (z,x)= \frac{1}{\sqrt{(z-{\sigma}_0(x))^2-4{\tilde{\rho}}_0(x)}}
\end{equation}
and 
\begin{eqnarray}
	\label{abcd}
	\nonumber & &\alpha_{k,1}(x)-2\tilde{{\rho}}_k(x),\quad  
	\beta_{k,1}(x)-{\sigma}_k(x),\\ 
	& & \gamma_{k,1}(x)-4\tilde{{\rho}}_0(x){\sigma}_k(x),\quad 
		\xi_{k,1}(x)-2\tilde{{\rho}}_k(x),\\ 
	\nonumber
	& & \alpha_{k,j}(x),\quad  \beta_{k,j}(x),\quad \gamma_{k,j}(x),\quad \xi_{k,j}(x)
	\quad (j=2,\dots,2k-1),
\end{eqnarray}
are polynomials in  $\tilde{{\rho}}_0,\dots,\tilde{{\rho}}_{k-1},{\sigma}_0,\dots,{\sigma}_{k-1}$ and their $x$-derivatives.

Now we substitute the expansions (\ref{exgo})--(\ref{exgn}) in the continuum limit of the string equations~(\ref{str01}),
\begin{equation}
	\label{str1a}
		\frac{1}{2\pi\rmi}\oint_{\gamma_{\infty}}W_0'(z) T(\epsilon,z,x)\rmd z
		+
		\sum_{i=1}^m \mu_i \left(T(\epsilon,q_i,x)-1\right)=2 x,
\end{equation}		
\begin{equation}
	\label{str2a}		
				\frac{1}{2\pi\rmi}\oint_{\gamma_{\infty}}W_0'(z) R(\epsilon,z,x)\rmd z
		+
		\sum_{i=1}^m \mu_i R(\epsilon,q_i,x)=0.
\end{equation} 
In order to analyze the equations that  the system~(\ref{str1a})--(\ref{str2a}) imply for the coefficients
$\tilde{{\rho}}_k$ and ${\sigma}_k$  it is useful to introduce  the function
\begin{equation}
	\label{G}
	G({\sigma}_0,\tilde{{\rho}}_0)
	=
	-\oint_{\gamma_{\infty}}\frac{dz}{2\pi\mathrm{i}}\frac{W_0'(z)}{w(z,x)} 
	- \sum_{i=1}^m\mu_i\left(\frac{1}{w(q_i,x)}+{\sigma}_0\right).
\end{equation}
Notice that for $k=0$ the system (\ref{str1a})--(\ref{str2a}) can be rewritten as
\begin{equation}
	\label{sysrs0F}
	\frac{\partial G}{\partial \sigma_0}=2x,
	\quad
	\frac{\partial G}{\partial \tilde{{\rho}}_0}=0.
\end{equation}
Moreover, substituting~(\ref{RTk})--(\ref{abcd}) into the system it is straightforward to prove that for $k\geq 1$
both quantities
\begin{equation}
	\label{sysrsk}
	\begin{array}{lll}
	\partial_{{\sigma}_0}^2G({\sigma}_0,\tilde{{\rho}}_0){\sigma}_k+\partial_{{\sigma}_0}\partial_{\tilde{{\rho}}_0}
		G({\sigma}_0,\tilde{{\rho}}_0)\tilde{{\rho}}_k,\\
	\partial_{{\sigma}_0}\partial_{\tilde{{\rho}}_0}G({\sigma}_0,\tilde{{\rho}}_0){\sigma}_k+\partial_{\tilde{{\rho}}_0}^2
		G({\sigma}_0,\tilde{{\rho}}_0)\tilde{{\rho}}_k,
	\end{array}
\end{equation}
are differential polynomials in  $\tilde{{\rho}}_0,\dots,\tilde{{\rho}}_{k-1}$, ${\sigma}_0,\dots,{\sigma}_{k-1}$,
and their $x$-derivatives. Therefore, if
\begin{equation}
	\label{regcond}
		\everymath{\displaystyle}
		\left|
		\begin{array}{cc}
			\partial_{{\sigma}_0}^2G({\sigma}_0,\tilde{{\rho}}_0)
			&
			\partial_{{\sigma}_0}\partial_{\tilde{{\rho}}_0}G({\sigma}_0,\tilde{{\rho}}_0)
			\\
			\partial_{{\sigma}_0}\partial_{\tilde{{\rho}}_0}G({\sigma}_0,\tilde{{\rho}}_0)
			&
			\partial_{\tilde{{\rho}}_0}^2G({\sigma}_0,\tilde{{\rho}}_0)
	\end{array}\right| \neq 0,
\end{equation}
then all the coefficients ${\sigma}_k$ and $\tilde{{\rho}}_k$ can be recursively obtained from the linear system~(\ref{sysrsk}).

Finally, it is easy to prove  that if $R(\epsilon,z,x)$, $T(\epsilon,z,x)$, $r(\epsilon,x)$ and $s(\epsilon,x)$, 
solve~(\ref{reccl1})--(\ref{reccl2}) and~(\ref{str1a})--(\ref{str2a}), then so do
\begin{equation}
	\everymath{\displaystyle}
	\begin{array}{lll}
	\hat{R}(\epsilon,z,x) = R(-\epsilon,z,x+\epsilon),
	&  &
	\hat{T}(\epsilon,z,x) = T(-\epsilon,z,x),
	\\
	\hat{r}(\epsilon,x) = r(-\epsilon,x),
	&  &
	\hat{s}(\epsilon,x) = s(-\epsilon,x+\epsilon).
	\end{array}
\end{equation}
Moreover, it is clear from our preceding analysis that there is only one  solution of~(\ref{reccl1})--(\ref{reccl2})
and~(\ref{str1})--(\ref{str2}) with the form (\ref{exgo})--(\ref{exgn}). Therefore, it must be
\begin{equation}
	\label{cons}
	\everymath{\displaystyle}\begin{array}{lll}
	R(\epsilon,z,x) = R(-\epsilon,z,x+\epsilon),
	&  &
	T(\epsilon,z,x)=T(-\epsilon,z,x),
	\\
	r(\epsilon,x) = r(-\epsilon,x),
	&  &
	s(\epsilon,x) = s(-\epsilon,x+\epsilon).
	\end{array}
\end{equation}
Consequently we have proved the existence of the expansions of the form~(\ref{exr}) and~(\ref{exs})
and that they satisfy~(\ref{cons}). Note also that the constraint~(\ref{cons}) for $s$ implies 
\begin{equation}
	\label{exrr}
	s(\epsilon,x) \approx \sum_{k=0}^{\infty} \epsilon^{2 k} \widetilde{{\sigma}}_k \left(x+\frac{\epsilon}{2}\right).                  
\end{equation}
A similar structure for the expansion of $R$ holds.

It should be noticed  that the recurrence coefficients $r(\epsilon,x)$ and $s(\epsilon,x)$ for Penner
models have the same type of expansions in $\epsilon$ as those rigorously proved for matrix models
with polynomial potentials~\cite{BL05,KU00}. 

A similar analysis can be applied  to $Z_2$-symmetric Penner models in the two-cut case to
prove the existence of the expansions~(\ref{r2})--(\ref{r3}): we perform the continuum limit of 
the string and resolvent equations~(\ref{stre1}) and~(\ref{recce1}), and  introduce two
expansions $U$ and $V$ of the form~\cite{ALA11}
\begin{equation}
	\label{uv}
	\fl
	T_{2 n+1}(\lambda)\sim U(\epsilon,x,\lambda) \approx \sum_{k=0}^\infty U_k(x,\lambda) \epsilon^{2k},
	\quad
     	T_{2 n}(\lambda)\sim V(\epsilon,x,\lambda) \approx \sum_{k=0}^\infty V_k(x,\lambda) \epsilon^{2k},
\end{equation}
where 
\begin{equation}
	\label{ex}
	x=\frac{2 n}{N},
	\quad
	\epsilon=\frac{1}{N}.
\end{equation}
Then  identification of the coefficients of $\epsilon^{2k}$ in the continuum limit of the resolvent identities~(\ref{resc2a})
leads to a system of recurrence relations for the coefficients $U_k$ and $V_k$. Moreover, from the form~(\ref{resc2a})
it follows that the expressions for the  coefficients $V_k$ are obtained from those for $U_k$ under the substitution
$(\alpha_i,\beta_i)\rightarrow (\beta_i,\alpha_i)$. Furthermore, one finds that  the coefficients $U_k$ can be written in the form
\begin{equation}
	\label{rkj}
  	U_k = \sum_{j=0}^{3k} \left( R_{k,j} +\lambda S_{k,j}\right) u^{2j+1},\quad  k\geq 0,
\end{equation}
where
\begin{equation}
	u(\lambda,x)=\frac{1}{\sqrt{ \left(\lambda-(\alpha_0+\beta_0)\right)^2-4\alpha_0\beta_0}}.
\end{equation}
and $R_{k,j}=R_{k,j}(\alpha_0,\beta_0,\ldots,\alpha_k,\beta_k)$ and
$S_{k,j}=S_{k,j}(\alpha_0,\beta_0,\ldots,\alpha_k,\beta_k)$ are polynomials in $\alpha_0,\beta_0,\ldots,\alpha_k,\beta_k$
and their $x$ derivatives.

The final  step is to introduce  the expansions~(\ref{uv}) in the continuum limit~(\ref{stre1v}) of the string equation.
Then it can be proved that identifying powers of $\epsilon^2$ in~(\ref{stre1v}) determines recursively
all the coefficients $\alpha_k$ and $\beta_k$.
%%%%%%%%%%%%%%%%%%%%%%%%%%%%%%%%%%%%%%%%%%%%%%%%%%%%%%%%%%%%%%%%%
%% REFERENCES %%%%%%%%%%%%%%%%%%%%%%%%%%%%%%%%%%%%%%%%%%%%%%%%%%%%%%%
%%%%%%%%%%%%%%%%%%%%%%%%%%%%%%%%%%%%%%%%%%%%%%%%%%%%%%%%%%%%%%%%%
\section*{References}
%%%%%%%%%%%%%%%%%%%%%%%%%%%%%%%%%%%%%%%%%%%%%%%%%%%%%%%%%%%%
\providecommand{\newblock}{}

%%%%%%%%%%%%%%%%%%%%%%%%%%%%%%%%%%%%%%%%%%%%%%%%%%%%%%%%%%%%
%%  THE END %%%%%%%%%%%%%%%%%%%%%%%%%%%%%%%%%%%%%%%%%%%%%%%%%%%%
%%%%%%%%%%%%%%%%%%%%%%%%%%%%%%%%%%%%%%%%%%%%%%%%%%%%%%%%%%%%
\end{document}